\documentclass[twocolumn,aps,superscriptaddress,longbibliography]{revtex4-1} 
\usepackage{bm}
\usepackage{amsmath}
\usepackage{amssymb}
\usepackage{times}
\usepackage[usenames,dvipsnames]{color}
\usepackage{graphicx}
\usepackage{dcolumn}
\usepackage{simplewick}
\usepackage{hyperref}

\hypersetup{
  colorlinks=false,
  colorlinks=true,
  citecolor=blue,
  linkcolor=blue,
  urlcolor=blue}

\begin{document}
\title{Fock-space perturbed relativistic coupled-cluster theory for electric 
       dipole polarizability of one-valence atomic systems: Application to Al and In}

\author{Ravi Kumar}
\affiliation{Department of Physics, Indian Institute of Technology, 
             Hauz Khas, New Delhi 110016, India}


\author{D. Angom}
\affiliation{Department of Physics, Manipur University,
             Canchipur 795003, Manipur, India}
\affiliation{Physical Research Laboratory,
             Ahmedabad - 380009, Gujarat,
             India}

\author{B. K. Mani}
\email{bkmani@physics.iitd.ac.in}
\affiliation{Department of Physics, Indian Institute of Technology, 
             Hauz Khas, New Delhi 110016, India}

\begin{abstract}

We have developed a Fock-space relativistic coupled-cluster theory based
method for the calculation of electric dipole polarizability of one-valence
atoms and ions. We employ this method to compute the ground-state and
spin-orbit coupled excited state electric dipole polarizability of Al and In.
To check the quality of many-electron wavefunctions, we also compute the excitation 
energies of some low-lying states of Al and In. The effects of Breit interaction 
and QED corrections from the Uehling potential and the 
self-energy are included to improve the accuracy of $\alpha$ further. 
Our recommended value of ground-state $\alpha$ for both atoms are in good 
agreement with the previous theoretical results. From our computations, we 
find that more than 65\% of contributions come from the dipolar mixing of 
$3p$($5p$) with $3d$($5d$) and $4s$($6s$)-electrons for Al(In). The largest 
Breit and QED contributions are found to be 1.3\% and 0.6\%, respectively. 
\end{abstract}

\maketitle

\section{Introduction}

Group-13 elements are a promising candidates for accurate optical atomic 
clocks as they are offer low fractional frequency 
errors \cite{dehmelt_82,chou-10,safronova-11b,zuhrianda-12,chen-17}. It is to 
be mentioned that, the recent experiments, 
Refs. \cite{brewer_19a,brewer_19b}, on Al$^+$ optical atomic clock have 
achieved a low fractional frequency error 
of $9.4\times10^{-19}$ \cite{brewer_19a,brewer_19b} which is, perhaps, the 
most accurate clock in existence today. The electric dipole polarizability, 
$\alpha$, of an atom or ion is a key parameter in estimating the accuracy of
an atomic clock. It is used to estimate the blackbody radiation (BBR) shift, 
one of the dominant environment induced frequency shifts, in the transition 
frequencies of atoms and ions due to ac Stark effect. 
Since the measurement of $\alpha$ for individual states is 
non-trivial \cite{wijngaarden-99}, accurate values from precision theory 
calculations play a crucial role in the development of new frequency and time 
standards for atomic clocks.
The other potential implications of $\alpha$ include, discrete symmetry 
violations in atoms and ions \cite{khriplovich-91, griffith-09}, condensates 
of dilute atomic gases \cite{anderson-95,bradley-95,davis-95}, high-harmonic 
generation and ultrafast processes \cite{lewenstein-94,lewenstein-95,blaga-09,lai-18}, 
and the search for the variation in the fundamental 
constants \cite{karshenboim-10, murphy-07}.

In this work, we have employed a Fock-space perturbed relativistic 
coupled-cluster (FS-PRCC) theory to compute the properties of one-valence
atomic systems in the external perturbations. We employ this method to compute 
the $\alpha$ for the ground state, $^2P_{1/2}$, and the spin-orbit (SO)-splitted 
excited state, $^2P_{3/2}$, of Al and In. In the literature, $\alpha$ for 
$^2P_{1/2}$ and $^2P_{3/2}$ states of Al and In have been calculated using 
different methods \cite{fleig-05,borschevsky-12,buchachenko-10,lupinetti-05,fuentealba-04,chu-04}, 
including the coupled-cluster based methods like ours. One common trend in the 
reported data is, however, a large variation in the $\alpha$ values. 
For example, for $^2P_{1/2}$ state of Al, there is a difference of 
$\approx$ 10\% in the smallest \cite{fleig-05} and the largest \cite{chu-04} 
reported $\alpha$ values.  The same trend is also observed in the experimental 
values \cite{milani-90,sarkisov-06,guella-84,lei-15}. The recent experiment
\cite{sarkisov-06} reports $\approx$ 20\% larger $\alpha$ than 
Ref. \cite{milani-90} for Al.
It is to be emphasized that, unlike the closed-shell atomic systems, 
calculation of $\alpha$ for an open-shell system is a challenging task and 
requires the inclusion of {\em core-core}, {\em core-valence} and 
{\em valence-valence} electrons correlations to the highest level of accuracy. 
Moreover, the inclusion of correlation effects from Breit interaction and QED 
corrections and the large basis sets are essential to tune the accuracy further.
The aim of the present work is to fill this gap. We aim to: develop a 
Fock-space relativistic coupled-cluster (FS-RCC) theory based method to 
accurately account for an external perturbation in the properties calculation 
of one-valence atomic systems; compute the accurate value of $\alpha$ for Al 
and In; and quantify the various electron correlation effects embedded in the
$\alpha$ of Al and In.

To test the accuracy of the wavefunctions, we have calculated to excitation 
energies of few low-lying states of Al and In using RCC theory. RCC is one of 
the most powerful many-body theories for atomic structure calculations.  
It accounts for the electron correlation to all-orders of residual Coulomb 
interaction, and has been used to calculate a plethora of properties in atomic 
systems. The implementation of such a theory and a FORTRAN code for the 
properties calculations of closed-shell and one-valence atomic systems without 
external perturbation is reported in our previous work \cite{mani-17}. 
For the properties calculation in the presence of external perturbation, we 
had reported a perturbed relativistic coupled-cluster (PRCC) theory 
for closed-shell in the works \cite{ravi-20, ravi-21} and references therein.
One of the key merits of PRCC is that it does not employ the 
sum-over-state \cite{safronova-99,derevianko-99} approach to incorporate the 
effects of a perturbation. The summation over all the possible intermediate 
states is subsumed in the perturbed cluster operators. 
Due to important prospects associated with $\alpha$, it has been computed 
using a variety of other many-body methods in the literature. The recent 
review article by Mitroy {\em et al.} \cite{mitroy-10} provides a summary 
of $\alpha$ for several atoms and ions computed using different methods. 
The other reference which we found very useful is the Schwerdtfeger's updated 
table of $\alpha$ for neutral atoms \cite{peter-19}. The table provides an 
exhaustive list of references on experimental and theoretical values 
of $\alpha$ for several neutral atoms.

The remaining part of the paper is organized into five sections. In Sec. II, 
we discussed the RCC and PRCC theories for one-valence atomic systems where 
we derive the PRCC equations and also discuss in detail the contributing
diagrams to each terms. In Sec. III, we discuss the calculation
of $\alpha$ using PRCC theory. Here, we provide some dominant diagrams 
contributing to $\alpha$. The basis set convergence and other calculational
details are discussed in Sec. IV of the paper. In Sec. V, we analyze and 
present our results of excitation energy and dipole polarizability. 
Unless stated otherwise, all the results and equations presented in this 
paper are in atomic units ($\hbar = m_e = e = 1/{4\pi\epsilon_0} = 1$).

\section{Methodology}

\subsection{One-valence RCC Theory}

The many-electron ground state wavefunction of an one-valence atom or ion in
the RCC theory is expressed as 
\begin{equation}
 |\Psi_v\rangle = e^{(T^{(0)}+S^{(0)})}\ |\Phi_v\rangle,
 \label{psi_1}
\end{equation} 
where $|\Phi_v\rangle$ is the one-valence Dirac-Fock (DF) reference state, 
and is obtained by adding an electron to the closed-shell
reference state, $|\Phi_v\rangle = a_v^ \dagger |\Phi_0\rangle$. The operators 
$T^{(0)}$ and $S^{(0)}$ are the coupled cluster (CC) operators which act 
within the Hilbert spaces of the closed-shell and open-shell systems, 
respectively. The ground state $|\Psi_v\rangle$ is the solution of 
the eigenvalue equation
\begin{eqnarray}
H^{\text{DCB}} |\Psi_v\rangle = E_v |\Psi_v\rangle,
   \label{psi_2}
\end{eqnarray}
where $H^{\text{DCB}}$ is the Dirac-Coulomb-Breit no-virtual-pair Hamiltonian 
and $E_v$ is the exact energy of the one-valence system. For an atom with 
$N$-electrons, $H^{\text{DCB}}$ is
\begin{eqnarray}
	H^{\text{DCB}} & = & \sum_{i=1}^N \left [c\bm{\alpha}_i \cdot
        \mathbf{p}_i + (\beta_i -1)c^2 - V_{N}(r_i) \right ]
                       \nonumber \\
   & & + \sum_{i<j}\left [ \frac{1}{r_{ij}}  + g^{\text{B}}(r_{ij}) \right ],
  \label{ham_dcb}
\end{eqnarray}
where $\bm{\alpha}$ and $\beta$ are the Dirac matrices, and $V_{N}(r_{i})$
is the nuclear potential. And, the last two terms, $1/r_{ij} $ and
$g^{\\text{B}}(r_{ij})$,  are the Coulomb and Breit interactions, respectively. 
The effects of the negative-energy continuum states are avoided by 
employing a kinetically balanced finite Gaussian 
basis \cite{mohanty-91,stanton-84}.

In the RCC theory, the single and double excitations incorporate most of the 
electron correlation effects and provide a good description of the properties. 
Therefore, we can approximate $T^{(0)} = T^{(0)}_1 + T^{(0)}_2$  and 
$S^{(0)} = S^{(0)}_1 + S^{(0)}_2$,  which is referred to as the coupled-cluster 
with singles and doubles (CCSD) approximation. These operators in the second 
quantized notation are 
\begin{subequations}
\begin{eqnarray}
   T^{(0)}_1  & = & \sum_{ap}t_a^p a_p^{\dagger}a_a, {\;\; \text{and} \;\;} 
   T^{(0)}_2  = \frac{1}{2!}\sum_{abpq}t_{ab}^{pq} 
                a_p^{\dagger}a_q^{\dagger}a_ba_a, \\
  S^{(0)}_1 & = &\sum_{p}s_v^p a_p^{\dagger}a_v,  {\;\; \text{and} \;\;}
  S^{(0)}_2 = \sum_{apq}s_{va}^{pq} a_p^{\dagger}a_q^{\dagger}a_aa_v.
\end{eqnarray}
\end{subequations}
Here, the indices $ab\ldots$ and $pq\ldots$ represent the core and virtual
orbitals, respectively. And, $t_{\cdots}^{\cdots}$ and $s_{\cdots}^{\cdots}$ 
are the cluster amplitudes of the $T$ and $S$ operators, respectively. These 
closed-shell and one-valence operators are obtained by solving a set of 
coupled nonlinear equations and details are discussed in our previous 
works \cite{mani-09,mani-10, mani-17}.  In the Ref. \cite{mani-17}, we have 
provided descriptions of the  computational implementation of RCC theory for 
the properties calculations of the closed-shell and one-valence systems 
without an external perturbation.


\subsection{One-valence PRCC theory}

In the presence of an external perturbation, the wavefunction and the energy 
of the system are modified. For the electric dipole polarizability, the 
perturbation is due to the interaction between the external electric field 
$\mathbf{E}_{\text{ext}}$ and the induced electric dipole moment of the system
$\mathbf{D}$. And, the interaction Hamiltonian is  
$H_1 = -\mathbf{D} \cdot \mathbf{E}_{\text{ext}}$. We refer the modified 
eigenstate as the perturbed eigenstate, $|\tilde{\Psi}_v \rangle$, and 
the modified energy as the perturbed energy, $\tilde{E}_v$. In the PRCC 
theory, $|\tilde{\Psi}_v \rangle$ is expressed as
\begin{eqnarray}
	|\tilde{\Psi}_v \rangle &=& e^{T^{(0)}} 
  \left[1 + \lambda \mathbf{T}^{(1)} \cdot \mathbf{E}_{\text{ext}}\right] 
       \nonumber \\ 
  &&  \left[1 + S^{(0)} + \lambda \mathbf{S}^{(1)} \cdot \mathbf{E}_{\text{ext}} 
     \right] |\Phi_v \rangle, 
 \label{ppsi_1}
\end{eqnarray} 
where, $\lambda$ is the perturbation parameter. The operators 
$\mathbf{T}^{(1)}$ and $\mathbf{S}^{(1)}$ are referred to as the perturbed 
closed-shell and one-valence cluster operators, respectively, and both are 
rank one operators. The operator $\mathbf{T}^{(1)}$ is obtained by solving a 
set of coupled perturbed equations within the Hilbert space of the occupied 
electrons. The details on its tensor representation and the PRCC equations 
are discussed in our previous works on the dipole polarizability of 
the closed-shell atomic systems \cite{chattopadhyay-12b,ravi-20}. So, here, 
we discuss only the tensor representation and PRCC equations of the 
open shell cluster operator $\mathbf{S}^{(1)}$.

\begin{figure}[h]
 \includegraphics[scale=0.5,angle=0]{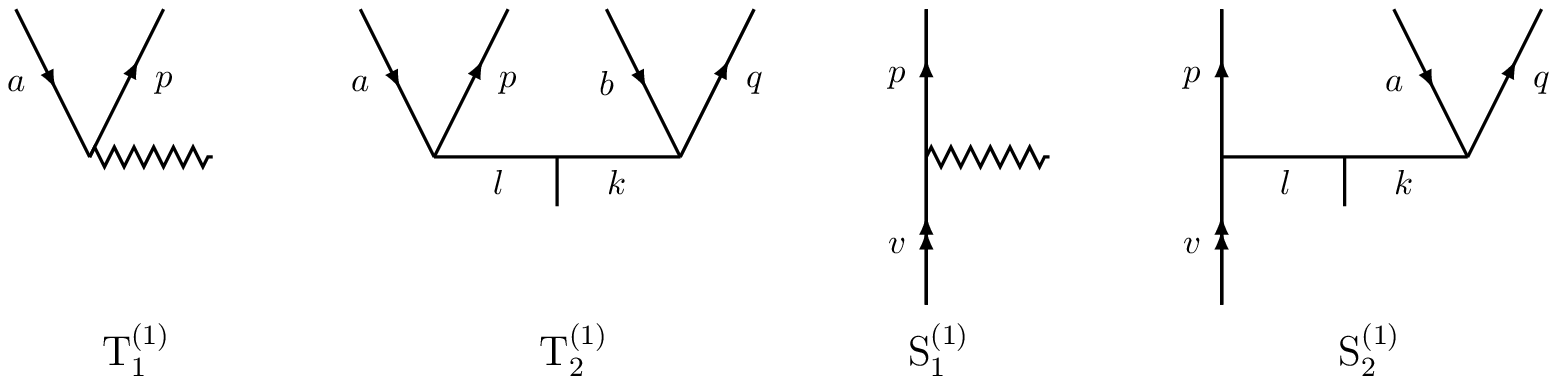}
	\caption{Diagrammatic representations of $\mathbf{T}_1^{(1)}$,
	$\mathbf{T}_2^{(1)}$, $\mathbf{S}_1^{(1)}$ and 
	$\mathbf{S}_2^{(1)}$ perturbed cluster operators.}
 \label{diag-ptrb}
\end{figure}

Similar to the case of $T^{(0)}$ and $S^{(0)}$ operators, in the CCSD 
approximation, we take $\mathbf{S}^{(1)} =  S_1^{(1)} + S_2^{(1)}$. And, 
these in the second quantized notations, 
\begin{subequations}
\begin{eqnarray}
   \mathbf{S}_1^{(1)} &= &\sum_{p}\xi_v^p \mathbf {C}_1(\hat r) 
            a_p^{\dagger}a_v, 
  \label{s1_1}    \\
  \mathbf{S}_2^{(1)}  &= &\sum_{apq} \sum_{lk} \xi_{va}^{pq}(l,k) 
  \mathbf{C}_l(\hat r_1)\mathbf{C}_k(\hat r_2) a_p^{\dagger}a_q^
  {\dagger}a_aa_v.  
  \label{S2_1}
\end{eqnarray}
\end{subequations}
Here, $\xi^{\cdots}_{\cdots}$ represents the cluster amplitude for the operator 
$\mathbf{S}^{(1)}$. The one-body operator $\mathbf{S}_1^{(1)}$ is an odd 
parity operator and expressed in terms of a rank-one $\mathbf{C}$-tensor. It 
satisfies the  orbital-parity and orbital-triangular selection rules, 
$(-1)^{l_v+l_p} = -1$ and  $|j_v-j_p| \leqslant 1 \leqslant (j_v + j_p)$, 
respectively. The tensor structure of the two-body operator 
$\mathbf{S}_2^{(1)}$ involves two $\mathbf{C}$-tensors with ranks $l$ and $k$ 
associated with its two-vertices. These two $\mathbf{C}$-tensors are coupled 
to give a rank-one operator, $\mathbf{S}_2^{(1)}$. The allowed orbital-parity 
and and orbital-triangular selections rules for $\mathbf{S}_2^{(1)}$ are, 
$(-1)^{l_v+l_p} = -(-1)^{l_a+l_q}$ and 
$|j_v-j_p| \leqslant l \leqslant (j_v+j_p)$, 
$|j_a-j_q| \leqslant k \leqslant (j_a+j_q)$, respectively. The diagrammatic 
representations of $\mathbf{T}^{(1)}$ and $\mathbf{S}^{(1)}$ are shown in 
the Fig.\ref{diag-ptrb}. 

In analogy with the Eq. (\ref{psi_2}), $|\tilde{\Psi}_v \rangle$ is the 
solution of the eigenvalue equation
\begin{equation}
  \left ( H^{\text{DCB}} + \lambda H_1 \right ) |\tilde{\Psi}_v \rangle 
   = \tilde{E}_v |\tilde{\Psi}_v \rangle,
\label{ppsi_2}
\end{equation}
here, within the first-order time-independent perturbation 
theory, the perturbed energy $\tilde{E}_v \equiv E_v$ as the first-order 
correction vanishes due to the odd parity nature of $H_1$. Using 
Eq. (\ref{ppsi_1}) in the eigenvalue equation (\ref{ppsi_2}) and by operating 
with $e^{-T^{(0)}}$ from left, retaining the terms first order in $\lambda$, 
we get
\begin{widetext}
\begin{eqnarray}
     &&\left [ e^{-T^{(0)}} H^{\text{DCB}} e^{T^{(0)}} \left(\mathbf{S}^{(1)} 
       \cdot \mathbf{E}_{\text{ext}}\right) + e^{-T^{(0)}} H^{\text{DCB}} 
       e^{T^{(0)}} \left(\mathbf{T}^{(1)} \cdot \mathbf{E}_{\text{ext}}\right) 
       \left(1 + S^{(0)}\right) + e^{-T^{(0)}} H_{1} e^{T^{(0)}} \left(1 
       + S^{(0)}\right) \right] |\Phi_v\rangle \\ \nonumber
     && = \left [ E_v \left ( \mathbf{S}^{(1)} \cdot \mathbf{E}_{\text{ext}} 
       \right ) + E_v \left (\mathbf{T}^{(1)} \cdot \mathbf{E}_{\text{ext}} 
       \right) \left (1 + S^{(0)} \right ) \right ]|\Phi_v\rangle.  
\label{peqn_1}
\end{eqnarray}
\end{widetext}
Using the definition of the normal ordered Hamiltonian, 
$H_N = H^{\text{DCB}} - \langle \Phi_v|H^{\text{DCB}}|\Phi_v\rangle$, and 
dropping $E_{\text{ext}}$ for simplicity from both sides of the equation, we 
can write
\begin{eqnarray}
  && \left [\bar{H}_N \mathbf{S}^{(1)} + \bar{H}_N \mathbf{T}^{(1)} 
     \left(1+S^{(0)}\right)+ \bar{H}_1 (1+S^{(0)}) \right ] |\Phi_v \rangle =  
        \nonumber \\
  && \Delta E_v \left [ \mathbf{S}^{(1)} + \mathbf{T}^{(1)} \left(1 
     + S^{(0)}\right) \right] |\Phi_v \rangle  
  \label{peqn_2}
\end{eqnarray}
where, $\Delta E_v,  = E_v - \langle \Phi_v|H^{\text{DCB}}|\Phi_v\rangle$ is 
the correlation energy of one-valence atom.  And, 
$\bar{H}_N, = e^{-T^{(0)}} H_N e^{T^{(0)}}$, is a similarity 
transformed Hamiltonian. Using the Wick's theorem, it can be reduced to
\begin{eqnarray}
  \bar{H}_N &=& H_N + \{\contraction[0.4ex]{} {H}{_N} {T} H_N T^{(0)}\} 
          + \frac{1}{2!} \{\contraction[0.4ex]{}{H}{_N}{T} 
         \contraction[0.8ex] {}{H}{_NT_0}{T^{(0)}}H_NT^{(0)}T^{(0)} \} 
                  \nonumber \\ 
           &&+ \frac{1}{3!}\{\contraction[0.4ex]{}{H}{_N}{T} 
          \contraction[0.8ex] {}{H}{_N T_0}{T^{(0)}}
          \contraction[1.2ex]{}{H}{_N T^{(0)} T_0}{T^{(0)}} 
          H_N T^{(0)} T^{(0)} T^{(0)} \} 
          + \frac{1}{4!}\{\contraction[0.4ex]{}{H}{_N}{T} 
          \contraction[0.8ex] {}{H}{_N T_0}{T^{(0)}}
          \contraction[1.2ex]{}{H}{_N T^{(0)} T_0}{T^{(0)}} 
          \contraction[1.6ex]{}{H}{_N T^{(0)}T^{(0)} T_0}{T^{(0)}} 
          H_N T^{(0)} T^{(0)} T^{(0)} T^{(0)} \} 
\label{hnbar}
\end{eqnarray}
By projecting Eq. (\ref{peqn_2}) with singly and doubly-excited determinants, 
$\langle \Phi_v^p|$ and $\langle \Phi_{va}^{pq}|$, respectively, and using 
the Wicks's theorem to remove the disconnected terms, we obtain the PRCC 
coupled equations for singles and doubles as
\begin{widetext}
\begin{subequations}
\begin{eqnarray}
  \langle \Phi_v^p |\bar H_1 +  \{\contraction[0.4ex] {\bar}{H}{_1}{S} 
      \bar H_1 S^{(0)}\} + \{\contraction[0.4ex] {\bar}{H}{_N}{T} 
      \contraction[0.8ex] {\bar}{H}{_N T^{(1)}(1+}{S} {\bar H}_N 
      \mathbf{T}^{(1)} (1 + S^{(0)})\} + \{\contraction[0.4ex] 
      {\bar}{H}{_N}{S} \bar H_N \mathbf{S}^{(1)}\} \Phi_v \rangle 
      & = & E_v^{\text{att}} \langle \Phi_v^p|\mathbf{S}_1^{(1)}| 
      \Phi_v \rangle,  
  \label{peqn_3}
\end{eqnarray}
\begin{eqnarray}
  \langle \Phi_{va}^{pq} |\bar H_1 +  \{\contraction[0.4ex] 
      {\bar}{H}{_1}{S} \bar H_1 S^{(0)}\} + \{\contraction[0.4ex] 
      {\bar}{H}{_N}{T} \contraction[0.8ex] {\bar}{H}{_N T^{(1)}(1+}{S} 
      {\bar H}_N \mathbf{T}^{(1)} (1+S^{(0)})\} + \{\contraction[0.4ex] 
      {\bar}{H}{_N}{S} \bar H_N \mathbf{S}^{(1)}\} \Phi_v \rangle 
      & = & E_v^{\text{att}} \langle \Phi_{va}^{pq}|\mathbf{S}_2^{(1)}
      |\Phi_v \rangle.
  \label{peqn_4}
\end{eqnarray}
\end{subequations}  
\end{widetext}
Here, $E_v^{\text{att}}$ is the attachment energy of the valence electron and
is expressed as $E_v^{\text{att}} = \epsilon_v + \Delta E_v$, where 
$\epsilon_v$ is the single-particle energy. In deriving these equations we have
used the relations, $\langle \Phi_v^{*}|\mathbf{T}^{(1)}|\Phi_v \rangle=0$
and $\langle\Phi_v^{*} |\mathbf{T}^{(1)} S^{(0)}|\Phi_v \rangle=0$, as they
do not contribute, where $*$ represents the single and doubly excited
determinant. We solve these coupled nonlinear equations using the 
Jacobi method. To remedy the slow convergence of this method we 
employ direct inversion of the iterated subspace (DIIS) \cite{pulay-80}.


\subsection{Linearized PRCC}

The Eqs. (\ref{peqn_3}) and (\ref{peqn_4}) contain all the CC terms 
associated with the PRCC equations of the one-valence system. And, therefore, 
provides an accurate description of the properties of the system. However, 
solving these equations is computationally expensive due to the large number 
of many-body diagrams arising from the contractions with multiple CC 
operators. One simple approach to mitigate this is to retain terms which are
linear in the CC operators. And, this also provides reliable results as in 
most of the cases the contribution from the nonlinear terms is small. So, 
retaining the terms linear in CC operators, we can write Eqs. (\ref{peqn_3}) 
and (\ref{peqn_4}) as
\begin{subequations}
\begin{eqnarray}
  && \langle \Phi_v^p |H_1 + \{\contraction[0.4ex] {}{H}{_1}{T} H_1 T^{(0)}\} 
     + \{\contraction[0.4ex] {}{H}{_1}{S} H_1 S^{(0)}\} 
     + \{\contraction[0.4ex] {}{H}{_N}{T} H_N \mathbf{T}^{(1)}\} 
       \nonumber \\
  && + \{\contraction[0.4ex] {}{H}{_N}{S} H_N \mathbf{S}^{(1)}\} \Phi_v \rangle 
     = E_v^{\text{att}} \langle \Phi_v^p|\mathbf{S}_1^{(1)}| \Phi_v \rangle,
  \label{peqn_5}   \\
  && \langle \Phi_{va}^{pq} |H_1 
     + \{\contraction[0.4ex] {}{H}{_1}{T} H_1 T^{(0)}\} 
     + \{\contraction[0.4ex] {}{H}{_1}{S} H_1 S^{(0)}\} 
     + \{\contraction[0.4ex] {}{H}{_N}{T} H_N \mathbf{T}^{(1)}\} 
       \nonumber \\
  && + \{\contraction[0.4ex] {}{H}{_N}{S} H_N \mathbf{S}^{(1)}\} \Phi_v \rangle 
     = E_v^{\text{att}}\langle \Phi_{va}^{pq}|\mathbf{S}_2^{(1)}| 
     \Phi_v \rangle.
  \label{peqn_6}
\end{eqnarray}
\end{subequations}
We refer to these equations as the linearized perturbed coupled-cluster 
(LPRCC) equations. The LPRCC incorporates all the important many-body effects 
like random-phase approximation and provides a good description of the 
one-valence atomic or ionic properties in the presence of perturbation.


\subsection{PRCC diagrams}

To solve the coupled-cluster amplitude Eqs.(\ref{peqn_3}) and (\ref{peqn_4}),
we have to evaluate all the possible terms arising from each of the matrix
elements in the equations. There are several terms and the book keeping is 
simplified with the many-body Goldstone diagrammatic approach. 
In this section we describe the Goldstone diagrams arising from the matrix
elements and these are evaluated manually. It is, however, possible to 
identify diagrams computationally as well \cite{fritzsche-08}. We have adopted
the manual approach for the convenience in evaluating the angular factors. For 
simplicity, from here onwards, as it should be the case for $\alpha$, 
we use $\mathbf{D}$ in place of $H_1$.


\subsubsection{$\bar{\mathbf{D}}$}

\begin{figure}
 \includegraphics[width=7.0cm, angle=0]{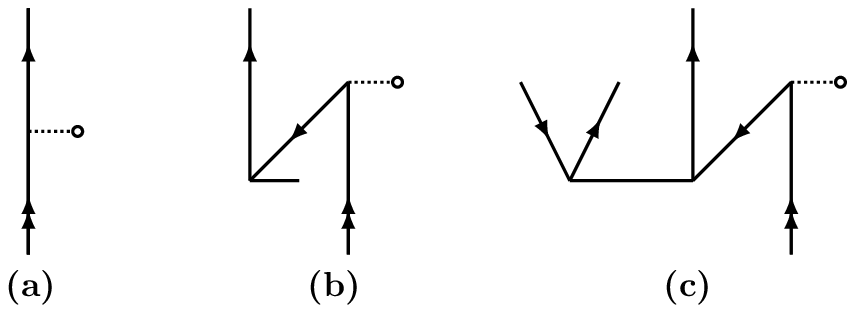}
 \caption{Single and double PRCC diagrams contributing to the term 
	 $\bar {\mathbf{D}}$ of Eqs. (\ref{peqn_3}) and (\ref{peqn_4}),
	 respectively.}
 \label{diag-t1}
\end{figure}

For the one-valence system, from the definition of the similarity
transformed Hamiltonian in Eq. (\ref{hnbar}), using the CSSD approximation 
we get
\begin{equation}
   \bar{\mathbf{D}} = \mathbf{D} 
   + \contraction[0.4ex]{ }{\mathbf{D}}{}{T}\mathbf{D}T_1^{(0)} 
   + \contraction[0.4ex]{}{\mathbf{D}}{}{T}\mathbf{D}T_2^{(0)}.
  \label{term1}
\end{equation}
The terms with two or higher orders of $T^{(0)}$ do not contribute
to the PRCC equation for one-valence system. The first term, $\mathbf{D}$, 
is the bare dipole operator and contributes to the equation of 
$\mathbf{S}_1^{(1)}$.  The two remaining terms represent the contraction with 
the unperturbed operator $T^{(0)}$ and contribute to the PRCC equations of 
the $\mathbf{S}^{(1)}_1$ and $\mathbf{S}^{(1)}_2$, respectively. In total, 
there are 3 diagrams from $\bar{\mathbf{D}}$, and these are shown in 
Fig. \ref{diag-t1}. Using the algebra of evaluating the Goldstone 
diagrams \cite{lindgren-86}, we can write the algebraic expressions 
of the diagrams as
\begin{subequations}
\begin{eqnarray}
   \langle  \mathbf{D} \rangle _v^p + \langle \contraction[0.4ex]{}{D}{}{T} 
       \mathbf{D}T^{(0)} \rangle ^p_v & = & \mathbf{r}_{pv} 
       - \sum_{a} \mathbf{r}_{av} t_a^p, 
		   \\
   \langle \contraction[0.4ex] { }{D}{}{T} \mathbf{D}T^{(0)} 
       \rangle _{va}^{pq} &=& - \sum_{b} \mathbf{r}_{bv} t_{ab}^{qp},
\label{H1bar-2b}
\end{eqnarray}
\end{subequations}
respectively. Here, in atomic units $\mathbf{D}\equiv-\mathbf{r}$ and 
$\mathbf{r}_{ij} = \langle i|\mathbf{r}|j \rangle$ is the electronic part of 
the single-particle matrix element of the dipole operator.  And, 
$\langle \cdots \rangle _v^p$ and $\langle \cdots \rangle _{vb}^{pq}$ 
represent the matrix elements $\langle \Phi_v^p| \cdots | \Phi_v \rangle$ and 
$\langle \Phi_{va}^{pq}| \cdots| \Phi_v \rangle$, respectively.


\subsubsection{ $\bar {\mathbf{D}} S^{(0)}$}

\begin{figure}
\includegraphics[width=7.5cm, angle=0]{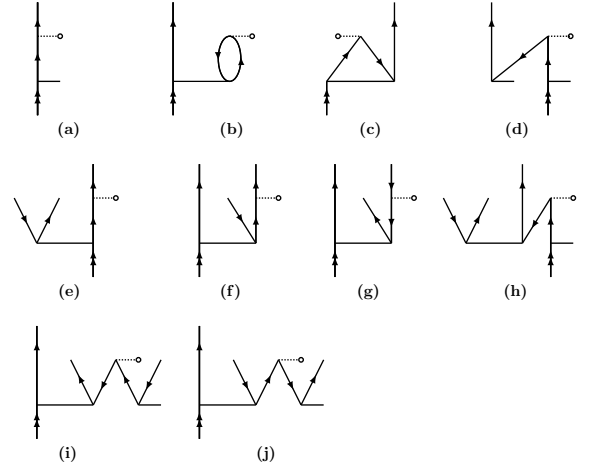}
\caption{Single and double PRCC diagrams contributing to the term
	$\contraction[0.4ex]{}{D}{}{S}\bar{\mathbf{D}}S^{(0)}$ of
	PRCC Eqs. (\ref{peqn_3}) and (\ref{peqn_4}),
	respectively.}
\label{diag-t2}
\end{figure}

Like the first term, consider the second term in Eqs.(\ref{peqn_3}) and 
(\ref{peqn_4}). Expanding the similarity transformed operator 
$\bar {\mathbf D}$ in terms of $T^{(0)}$, we can write 
\begin{eqnarray}
  \contraction[0.4ex]{}{D}{}{S} {\bar{\mathbf{D}}} S^{(0)} &=& 
  \contraction[0.4ex]{}{D}{}{S} \mathbf{D} S^{(0)}_1 +
  \contraction[0.4ex]{}{D}{}{S} \mathbf{D} S^{(0)}_2 +
  \contraction[0.4ex]{}{D}{}{T} \contraction[0.8ex]{}{D}{T^{(0)}}{S}
  \mathbf{D}T^{(0)}_1 S^{(0)}_1 + 
  \contraction[0.4ex]{}{D}{}{T} \contraction[0.8ex]{}{D}{T^{(0)}}{S}
  \mathbf{D}T^{(0)}_2S^{(0)}_1 \nonumber \\ 
 &&+ \contraction[0.4ex]{}{D}{}{T} \contraction[0.8ex]{}{D}{T^{(0)}}{S}
  \mathbf{D}T^{(0)}_1S^{(0)}_2.
\end{eqnarray}
The terms having higher than two orders of CC operators do not contribute. The 
first three terms in the above equation contribute to the PRCC equation of
$\mathbf{S}^{(1)}_1$ and leads to 4 diagrams. The diagrams are shown 
in Fig. \ref{diag-t2}(a-d). Except for the first term, all the other terms 
contribute to the PRCC equation of $\mathbf{S}^{(1)}_2$. In total there are 
6 diagrams from these terms and these are shown in Fig. \ref{diag-t2}(e-j). 
The corresponding algebraic expressions are
\begin{subequations}
\begin{eqnarray}
  && \langle \contraction[0.4ex]{}{D}{}{S}\mathbf{D}S^{(0)} \rangle_v^p 
     + \langle \contraction[0.4ex]{}{D}{}{T} 
     \contraction[0.8ex]{}{D}{T^{(0)}}{S} \mathbf{D}T^{(0)}S^{(0)} 
     \rangle _v^p = \sum_{q} \mathbf{r}_{pq} s_v^q 
     + \sum_{aq} \mathbf{r}_{aq} (s_{va}^{pq} \nonumber \\ 
  &&\;\;\;- s_{va}^{qp} - t_a^p s_v^q ), \\
  && \langle \contraction[0.4ex] {}{D}{}{S} \mathbf{D}S^{(0)} \rangle_{va}^{pq} 
     + \langle \contraction[0.4ex]{}{D}{}{T} 
     \contraction[0.8ex]{}{D}{T^{(0)}}{S} \mathbf{D}T^{(0)}S^{(0)} 
     \rangle_{va}^{pq}= \sum_{r} (\mathbf{r}_{pr} s_{av}^{qr} 
     + \mathbf{r}_{qr} s_{va}^{pr}) \nonumber \\
  &&\;\;\;- \sum_b \mathbf{r}_{ba} s_{vb}^{pq}-\sum_{br}
    \mathbf{r}_{br}(t_{ab}^{qp} s_{v}^{r} + t_{a}^{r} s_{vb}^{pq} 
    + t_{b}^{q} s_{va}^{pr} ).
\end{eqnarray}
\end{subequations}
This term is an important one in PRCC theory as it subsumes the many-body
effects of the core-polarization. 

\subsubsection{${\bar H}_N \mathbf{T}^{(1)}$}

\begin{figure}
 \includegraphics[width=7.5cm, angle=0]{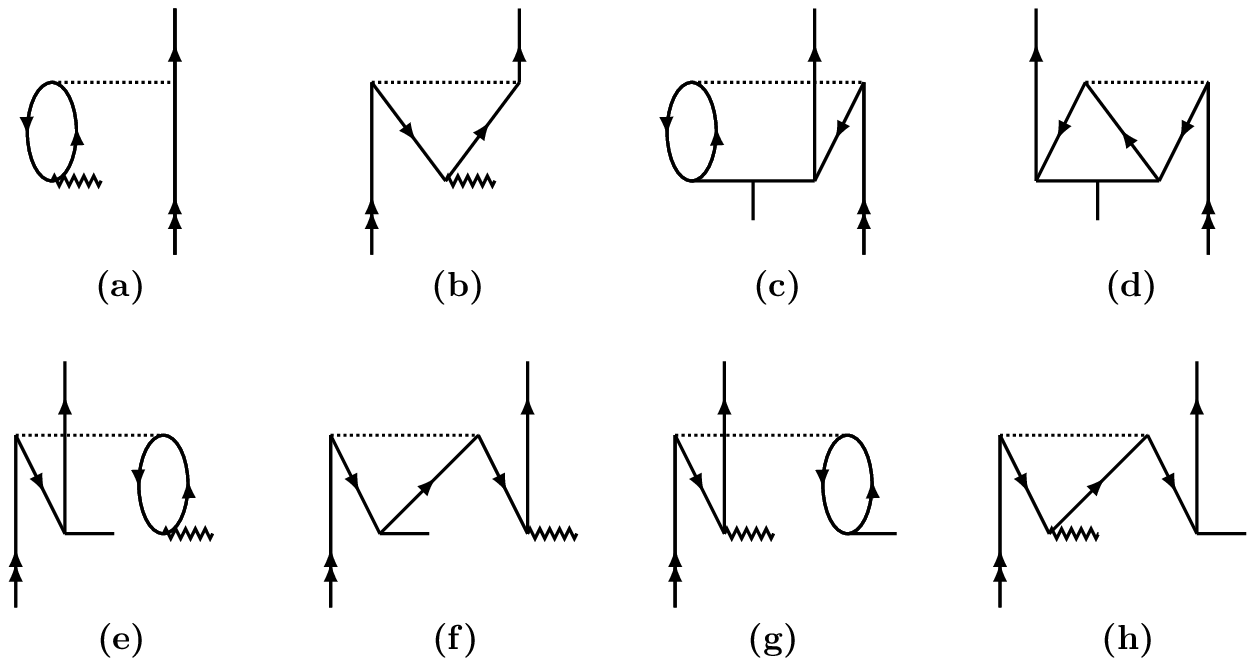}
\caption{Single PRCC diagrams contributing to the term 
	$\contraction[0.4ex]{\bar}{H}{_N}{T} {\bar H}_N \mathbf{T}^{(1)}$ of
Eq. (\ref{peqn_3}).}
 \label{diag-t3s1}
\end{figure}

Unlike the previous two terms where the dipole operator appears 
explicitly, in this term, the effects of the perturbation is embedded in
a rank one operator, $\mathbf{T}^{(1)}$. We can expand this as
\begin{equation}
  \!\!\!\! \contraction[0.4ex]{\bar }{H}{_N}{T} {\bar H}_N \mathbf{T}^{(1)} = 
   \contraction[0.4ex]{}{H}{_N}{T}H_N\mathbf{T}^{(1)} \! 
 + \contraction[0.4ex]{}{H}{_N}{T} \contraction[0.8ex]
    {}{H}{_NT^{(0)}}{T}H_N T^{(0)}\mathbf{T}^{(1)} \!
 + \frac{1}{2}\contraction[0.4ex]{}{H}{_N}{T} 
   \contraction[0.8ex]{}{H}{_N T^{(0)}}{T} 
   \contraction[1.2ex]{}{H}{_NT^{(0)}T^{(0)}}{T} H_NT^{(0)}T^{(0)}
   \mathbf{T}^{(1)}  
\label{t3}
\end{equation}
Consider first the PRCC equation of the $\mathbf{S}_1^{(1)}$. Only the first 
two terms contribute. The PRCC diagrams from these terms are obtained by 
invoking all the possible contractions between the $H_N$ and CC operators. 
There are  8 diagrams and these are shown in Fig. {\ref{diag-t3s1}}. 
We do not consider the diagrams arising from the one-body part of $H_N$. 
These do not contribute as we use Dirac-Fock orbitals in our calculations. 
The algebraic expression of these diagrams are given in Eq.(\ref{t3s1}).
In the equation, $g_{ijkl}$ represents the matrix element 
$\langle ij |1/r_{12} + g^B(r_{12})|kl \rangle$ and 
$\tilde {g}_{ijkl}, = (g_{ijkl} - g_{ijlk}$), is an antisymmetric matrix 
element. 

\begin{figure*}
 \includegraphics[width=15.0cm, angle=0]{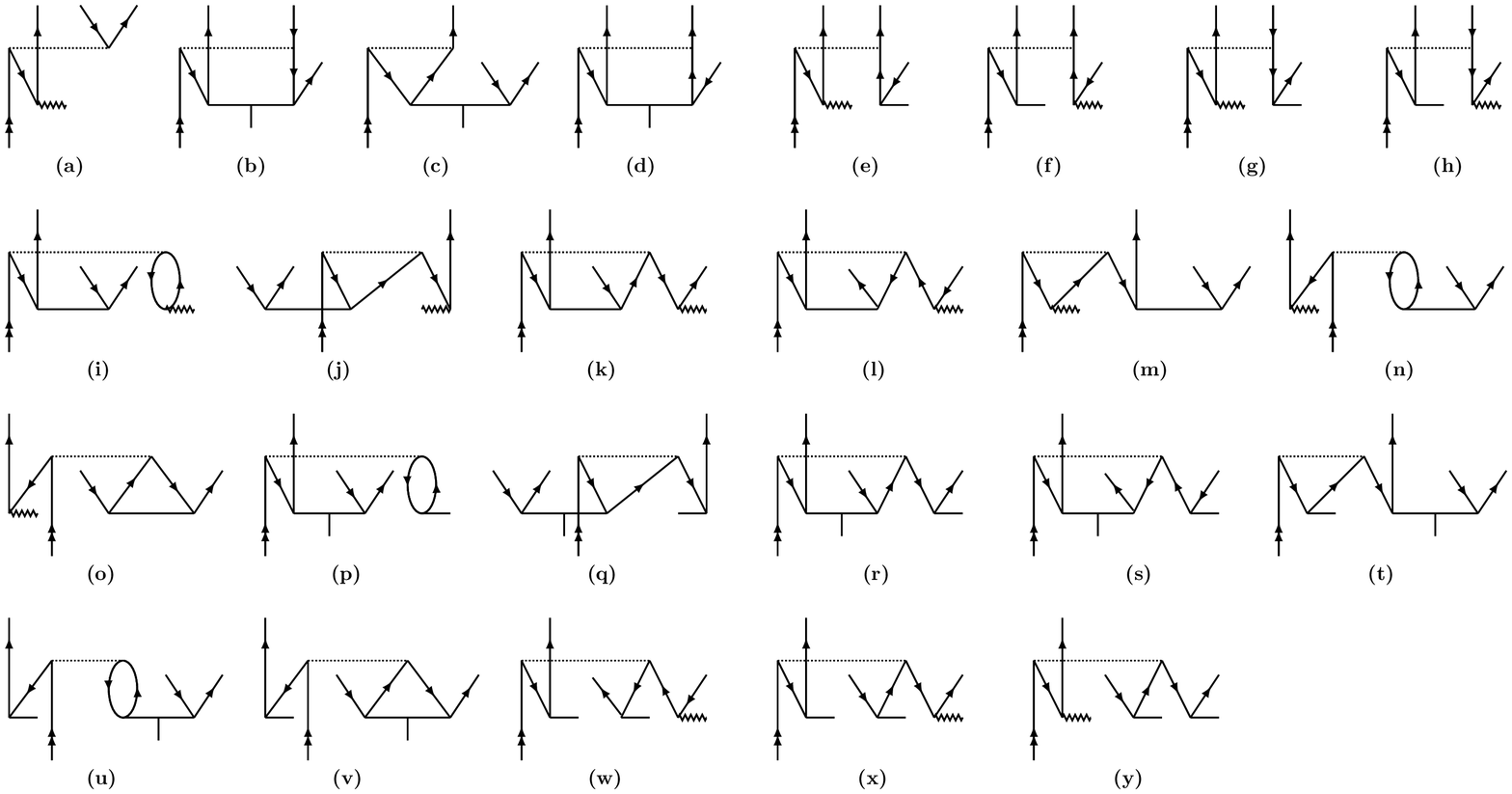}
 \caption{Double PRCC diagrams contributing to the term
	$\contraction[0.4ex]{\bar}{H}{_N}{T} {\bar H}_N \mathbf{T}^{(1)}$ of
        Eq. (\ref{peqn_4}).}
 \label{diag-t3s2}
\end{figure*}

In the PRCC equation of $\mathbf{S}_2^{(1)}$, all the terms contribute 
and leads to 29 diagrams. These diagrams are shown in Fig. \ref{diag-t3s2}. 
Like in the case of $\mathbf{S}^{(1)}$, we do not include the diagrams from 
the one-body part of the $H_N$. The algebraic expression of these diagrams 
are given in the Eq.(\ref{t3s2}). In the equation we see the emergence of 
a trend. The number of terms in this equation far exceed those in 
$\mathbf{S}_1^{(1)}$. This trend is there in the remaining non-linear 
terms as well.
\begin{widetext}
\begin{subequations}
\begin{eqnarray}
  && \langle \contraction[0.4ex]{}{H}{_N}{T} H_N\mathbf{T}^{(1)} \rangle _v^p 
    + \langle\contraction[0.4ex]{}{H}{_N}{T} 
    \contraction[0.8ex]{}{H}{_N T^{(0)}}{T}
    H_N T^{(0)}\mathbf{T}^{(1)} \rangle_v^p = 
    \sum_{aq} \tilde {g}_{apqv}\tau_a^q 
    + \sum_{abq} \left[- \tilde {g}_{abqv} \tau_{ab}^{qp} 
    +  g_{abvq} \left(- t_a^p\tau_b^q + t_a^q \tau_b^p 
    -  t_b^q \tau_a^p  + t_b^p \tau_a^q \right) \right] \\
\label{t3s1}
  && \langle \contraction[0.4ex]{}{H}{_N}{T} H_N\mathbf{T}^{(1)} 
    \rangle _{va}^{pq} + \langle \contraction[0.4ex]{}{H}{_N}{T} 
    \contraction[0.8ex] {}{H}{_NT^{(0)}} {T} H_NT^{(0)}\mathbf{T}^{(1)} 
    \rangle _{va}^{pq} + \langle \contraction[0.4ex]
   {}{H}{_N}{T} \contraction[0.8ex] {}{H}{_NT^{(0)}}{T}  \contraction[1.2ex]
   {}{H}{_NT^{(0)}T^{(0)}}{T} H_NT^{(0)}T^{(0)}\mathbf{T}^{(1)} 
   \rangle_{va}^{pq} =
   - \sum_{b} g_{bqva} \tau_b^p + \sum_{bc} g_{bcva} \tau_{bc}^{pq} 
   - \sum_{br} \left[g_{bpvr} \tau_{ba}^{rq} + g_{bqvr} (\tau_{ba}^{pr} 
    \right. \nonumber \\  
 &&\;\;+ \left. \tau_{b}^{p} t_{a}^{r} + t_{b}^{p} \tau_{a}^{r} ) \right] 
   + \sum_{bc} g_{bcva} (\tau_{b}^{p} t_{c}^{q} + t_{b}^{q} \tau_{c}^{q}) 
   + \sum_{bcr} g_{bcvr} \left( - t_{ba}^{pq} \tau_c^r + t_{ab}^{qr} \tau_c^p 
   + t_{ba}^{pr} \tau_c^q + t_{bc}^{pq} \tau_a^r +  t_{ca}^{pq} \tau_b^r 
   - t_{ca}^{rq} \tau_b^p + t_{ac}^{rq} \tau_b^p 
    \right. \nonumber \\ 
 &&\;\;- \left. \tau_{ba}^{pq} t_c^r + \tau_{ab}^{qr} t_c^p 
   + \tau_{ba}^{pr} t_c^q  + \tau_{bc}^{pq} t_a^r + \tau_{ca}^{pq} t_b^r 
   - \tau_{ca}^{rq} t_b^p + \tau_{ac}^{rq} t_b^p + \tau_{a}^{r} t_b^p t_c^q 
   + \tau_{c}^{q} t_b^p t_b^q + \tau_{b}^{p} t_a^r \right)
\label{t3s2}
\end{eqnarray}
\end{subequations}
\end{widetext}

\subsubsection{ ${\bar H}_N \mathbf{T}^{(1)} S^{(0)}$}

This term involves one each of the perturbed and unperturbed CC operators
$\mathbf{T}^{(1)}$ and $S^{(0)}$, respectively. And, the term contributes 
to the nonlinear PRCC equation. By expanding $\bar {H}_N$, we obtain
\begin{eqnarray}
  \contraction[0.4ex]{}{H}{_N}{T} \contraction[0.8ex]
  {}{H}{_NT^{(1)}}{S} {\bar H}_N\mathbf{T}^{(1)}S^{(0)} = \contraction[0.4ex]
  {}{H}{_N}{T} \contraction[0.8ex]{}{H}{_NT^{(1)}}{S} 
  H_N\mathbf{T}^{(1)}S^{(0)} + \contraction[0.4ex]{}{H}{_N}{T}
  \contraction[0.8ex] {}{H}{_NT^{(0)}}{T} 
  \contraction[1.2ex]{}{H}{_NT^{(0)}T^{(1)}}{S} H_NT^{(0)}
  \mathbf{T}^{(1)}S^{(0)}.    
\label{t4s2}
\end{eqnarray}
The higher order terms will not contribute. As to be expected, several
diagrams arise from this term due to various possible contractions 
between $H_{\rm N}$ and CC operators. For the PRCC equation of
$\mathbf{S}_1^{(1)}$, there are 14 diagrams from this term. And, these
are are shown in the left panel of Fig. \ref{diag-t4t5s1}. The algebraic 
expressions of these diagrams are given in Eq.(\ref{HNbar-T1S0_1b}).
\begin{widetext}
\begin{eqnarray}
   \langle \contraction[0.4ex]{}{H}{_N}{T} \contraction[0.8ex]
   {}{H}{_NT^{(1)}}{S} &H_N&\mathbf{T}^{(1)}S^{(0)} \rangle _v^p +
   \langle \contraction[0.4ex]{}{H}{_N}{T} \contraction[0.8ex]
   {}{H}{_NT^{(0)}}{T} \contraction[1.2ex] {}{H}{_NT^{(0)}T^{(1)}}{S}
   H_NT^{(0)}\mathbf{T}^{(1)}S^{(0)} \rangle _v^p = \sum_{aqr} g_{apqr} (\tau_a^q s_v^r
 - \tau_a^r s_v^q) + \sum_{aqbr} g_{abqr} \left[ s_v^r (- \tau_{ab}^{qp}
 +  \tau_{ab}^{pq}) + \tau_{a}^{q} s_{bv}^{rp} + \tau_{b}^{q} s_{av}^{rp}
 - \tau_{a}^{q} s_{bv}^{pr}  \right. \nonumber \\
 &+& \left. \tau_{a}^{p} s_{bv}^{qr} - \tau_{b}^{p}s_{av}^{qr}
 + \tau_{b}^{q} s_{av}^{pr} + (-\tau_{b}^{p} t_a^q - \tau_{a}^{q} t_b^p
 + \tau_{b}^{q} t_a^p + \tau_{a}^{p} t_b^q) s_{v}^{r} \right] 
\label{HNbar-T1S0_1b}
\end{eqnarray}
\begin{eqnarray}
  \langle \contraction[0.4ex]{}{H}{_N}{T} \contraction[0.8ex]
   {}{H}{_NT^{(1)}}{S} &H_N&\mathbf{T}^{(1)}S^{(0)} \rangle _{va}^{pq} +  
   \langle \contraction[0.4ex]{}{H}{_N}{T} \contraction[0.8ex]
   {}{H}{_NT^{(0)}}{T}\contraction[1.2ex] {}{H}{_NT^{(0)}T^{(1)}}{S}
   H_NT^{(0)}\mathbf{T}^{(1)}S^{(0)} \rangle_{va}^{pq} = 
   \sum_{rs} g_{pqrs} \tau_a^s s_v^r - \sum_{rb} ( g_{pbra} \tau_b^q + g_{qbar} \tau_b^p ) s_v^r 
 + \sum_{sbr} g_{pbsr} \left[s_v^s ( \tau_{ba}^{rq}
 - \tau_{ab}^{rq} ) - s_{v}^{r} (\tau_{ab}^{qs}  \right. \nonumber \\
 &+& \left.  \tau_{ab}^{sp} ) \right] +
   \sum_{cbr} g_{cbar} \tau_{cb}^{qp} s_v^r
 + \sum_{brs} g_{bprs}(s_{va}^{sq} \tau_b^r - s_{va}^{rq} \tau_b^s - s_{bv}^{qs} \tau_a^r 
 - s_{av}^{rs} \tau_b^q - s_{bv}^{pr} \tau_a^s - s_{bv}^{sp} \tau_a^r + s_{sp}^{av} \tau_b^r
 - s_{rs}^{va} \tau_b^p + s_{bv}^{rp} \tau_a^s \nonumber \\ 
 &-& s_{av}^{rp} \tau_b^s ) 
 + \sum_{cbr} g_{cbar} [ s_{bv}^{pr} \tau_c^q - s_{cv}^{qp} \tau_b^r
 +  s_{cv}^{rp} \tau_b^q + s_{bv}^{qp} \tau_c^r - s_{bv}^{rp} \tau_c^q 
 + s_{vc}^{rq} \tau_b^p ] + \sum_{bcrs} g_{bcrs} \left [s_{cv}^{sp} \tau_{ab}^{qr} 
 + s_{cv}^{rp} \tau_{ab}^{qs} - s_{cv}^{sp} \tau_{ab}^{rq} + s_{cv}^{qp} \tau_{ab}^{rs} \right. \nonumber \\
 &+& \left. s_{av}^{sp} \tau_{bc} ^{qr}   
 - s_{cv}^{ps} \tau_{ab}^{qr} + s_{cv}^{rs} \tau_{ab}^{qp}
 + s_{vc}^{rp} \tau_{ab}^{qs} + s_{vc}^{sp} \tau_{ab}^{rq} 
 - s_{vc}^{rs} \tau_{ab}^{qp} - s_{av}^{sp}
\tau_{bc}^{rq} - s_{cv}^{qp} \tau_{ba}^{rs} \right] 
\label{HNbar-T1S0_2b-1}
\end{eqnarray}
\begin{eqnarray}
   \langle \contraction[0.4ex]{}{H}{_N}{T} \contraction[0.8ex]{}{H}{_NT^{(0)}}{T}
   \contraction[1.2ex]{}{H}{_N T^{(0)} T^{(1)}}{S} &H_N& T^{(0)} \mathbf{T}^{(1)}S^{(0)} \rangle _{va}^{pq} + 
   \langle \contraction[0.4ex]{}{H}{_N}{T} \contraction[0.8ex]{}{H}{_NT^{(0)}}{T}
   \contraction[1.2ex]{}{H}{_NT^{(0)}T^{(1)}}{S} H_NT^{(0)}\mathbf{T}^{(1)}S^{(0)} \rangle_{va}^{pq} = 
   \sum_{brs} g_{bprs} \left [ (- t_a^r \tau_b^q - t_b^q \tau_a^r )s_v^s - (t_a^s \tau_b^p 
  + t_b^p \tau_a^s )s_v^r \right ] + \sum_{brc} g_{bcar} (t_b^q \tau_c^p \nonumber \\
 &+& t_c^p \tau_b^q )s_v^r +\sum_{brcs} g_{bcrs} \left [ (- t_{ab}^{qr} \tau_c^p + t_{ab}^{qp} \tau_c^r)s_v^s 
  + (- t_{ab}^{qp} \tau_c^s + t_{ab}^{qs} \tau_c^p) s_v^r + (t_{ab}^{rq} \tau_c^p 
  + t_{ac}^{rp} \tau_b^q + t_{bc}^{qp} \tau_a^r) s_v^s 
  + ( \tau_{ab}^{qs} t_c^p - \tau_{ab}^{qr} t_c^p \right. \nonumber \\
 &+& \left. \tau_{ab}^{qp} t_c^r - \tau_{ac}^{qp} t_b^r  
  + \tau_{ac}^{rp} t_b^q + \tau_{bc}^{qp} t_a^r + \tau_{ab}^{rq} t_c^p) s_v^s
  + [-t_b^r \tau_c^p - t_c^p \tau_b^r + t_b^p \tau_c^r + \tau_b^p t_c^r] s_{va}^{sq}
  + (-t_b^q \tau_a^r - \tau_b^q t_a^r) s_{cv}^{sp} + (t_c^r \tau_b^q  \right. \nonumber \\
 &+& \left. \tau_c^r t_b^q) s_{av}^{sp} + (t_b^s \tau_a^r + \tau_b^s t_a^r) s_{cv}^{qp} 
  + (t_a^r \tau_b^q + \tau_a^r t_b^q) s_{cv}^{ps} + (t_a^r \tau_c^p + \tau_a^r
  t_c^p) s_{bv}^{qs} + (t_b^q \tau_c^p + \tau_b^q t_c^p) s_{av}^{rs} \right ]
\label{HNbar-T1S0_2b-2}
\end{eqnarray}
\end{widetext}
In the case of the PRCC equation for $\mathbf{S}_2^{(1)}$, there are 72 
Goldstone diagrams arise from this term. Out of these, 36 diagrams 
each arise from the first and second terms. These diagrams are shown 
in the left and right panels, respectively, of the Fig. \ref{diag-t4s2}. 
The algebraic expression of the diagrams in the left panel is given
in Eq.(\ref{HNbar-T1S0_2b-1}). Similarly, the algebraic expression of the 
diagrams in the right panel is given in Eq.(\ref{HNbar-T1S0_2b-2}).

\begin{figure*}
 \includegraphics[width=15.5cm,angle=0]{HN_bar_T1S0_HN_S1.eps}
  \caption{Single PRCC diagrams contributing to the terms 
   $\contraction[0.4ex]{\bar}{H}{_N}{T}\contraction[0.8ex] {}{H}{_NT^{(1)}}{S}
   {\bar H}_N\mathbf{T}^{(1)}S^{(0)}$ (panel (a)) and 
   $\contraction[0.4ex]{\bar}{H}{_N}{S}{\bar H}_N\mathbf{S}^{(1)}$ (panel (b)) 
   of Eq. (\ref{peqn_3}).}
 \label{diag-t4t5s1}
\end{figure*}

\begin{figure*}
   \includegraphics[width=16.5 cm, angle=0]{HN_bar_T1_S0.eps}
    \caption{Double PRCC diagrams contributing to the terms 
     $\contraction[0.4ex]{}{H}{_N}{T}\contraction[0.8ex]{}{H}{_NT^{(1)}}{S} 
     H_N\mathbf{T}^{(1)}S^{(0)}$ (left panel) and $\contraction[0.4ex]{}{H}{_N}{T} 
     \contraction[0.8ex]{}{H}{_NT^{(0)}}{T}\contraction[1.2ex]{}{H}{_NT^{(0)}T^{(1)}}{S} 
     H_NT^{(0)}\mathbf{T}^{(1)}S^{(0)}$ (right panel) of Eq. (\ref{peqn_4}).}
    \label{diag-t4s2}
\end{figure*}


\subsubsection{ ${\bar H}_N \mathbf{S}^{(1)}$}

This term contains a perturbed operator, $\mathbf{S}^{(1)}$, which 
subsumes dominant effects of perturbation for one-valence 
atomic systems. Expanding ${\bar H}_N$
\begin{equation}
   \contraction[0.4ex]{\bar}{H}{_N}{S} {\bar H}_N\mathbf{S}^{(1)} = 
   \contraction[0.4ex]{}{H}{_N}{S} H_N\mathbf{S}^{(1)}  
 + \contraction[0.4ex]{}{H}{_N}{T}\contraction[0.8ex]{}{H}{_NT^{(0)}}{S}
   H_NT^{(0)}\mathbf{S}^{(1)} 
 + \frac{1}{2!} \contraction[0.4ex]{}{H}{_N}{T} \contraction[0.8ex]
   {}{H}{_NT^{(0)}}{T}\contraction[1.2ex] {}{H}{_NT^{(0)}T^{(0)}}{S} 
   H_NT^{(0)}T^{(0)}\mathbf{S}^{(1)}    
\label{}
\end{equation}
In the PRCC equation of $\mathbf{S}^{(1)}_1$, both the one- and two-body 
CC operators from the first and second terms contribute. From the third term, 
however, only the term $H_N T_1^{(0)} T_1^{(0)} S_1^{(1)}$ contributes. 
There are 15 diagrams with contribute to the PRCC equation of 
$\mathbf{S}^{(1)}_1$, and these are shown in the right panel of 
Fig. \ref{diag-t4t5s1}. The algebraic expression of these diagrams is given
in the Eq.(\ref{HNbar-S1_1b}).

For the PRCC equation of $\mathbf{S}^{(1)}_2$, there are in total 59 diagrams. 
Out of these 29 diagrams arise from the first two terms. And, these diagrams 
are shown in the left panel of the Fig. \ref{diag-t5s2}. The remaining 30 
diagrams arise from the third term, and are shown in the right panel of 
the Fig. \ref{diag-t5s2}. The algebraic expression of diagrams in the left 
and right panels are given in Eqs. (\ref{HNbar-S1_2b-1}) and 
(\ref{HNbar-S1_2b-2}), respectively. 

\begin{widetext}
\begin{eqnarray}
  \langle \contraction[0.4ex]{}{H}{_N}{S} &H_N&\mathbf{S}^{(1)} \rangle _v^p + 
  \langle \contraction[0.4ex]{}{H}{_N}{T}\contraction[0.8ex] {}{H}{_NT^{(0)}}{S}
  H_NT^{(0)}\mathbf{S}^{(1)} \rangle _v^p 
 +\contraction[0.4ex]{}{H}{_N}{T} \contraction[0.8ex]
  {}{H}{_NT^{(0)}}{T}\contraction[1.2ex] {}{H}{_NT^{(0)}T^{(0)}}{S} 
   H_NT^{(0)}T^{(0)}\mathbf{S}^{(1)} \rangle _v^p 
 = \sum_{aqr} g_{apqr}  ( \xi_{av}^{qr} - \xi_{av}^{rq} 
 + t_a^q \xi_v^r - t_a^r \xi_v^q) + \sum_{aqbr} g_{abqr} [ (- t_{ab}^{qp} \nonumber \\ 
 &+& t_{ab}^{pq})\xi_v^r + t_{a}^{q} \xi_{bv}^{rp} 
 + t_{b}^{q} \xi_{av}^{rp} - t_{a}^{q} \xi_{bv}^{pr} + t_{a}^{p} \xi_{bv}^{qr} 
 - t_{b}^{p} \xi_{av}^{qr} 
 + t_{b}^{q} \xi_{av}^{pr}  
 + \frac{1}{2!}(- t_a^q t_b^p + t_a^p  t_b^q) \xi_v^r + \sum_q g_q^p \xi_v^q 
\label{HNbar-S1_1b}
\end{eqnarray}
\begin{eqnarray}
   \langle \contraction[0.4ex]{}{H}{_N}{S} &H_N&\mathbf{S}^{(1)} \rangle _{va}^{pq} +
   \langle \contraction[0.4ex]{}{H}{_N}{T}\contraction[0.8ex]{}{H}{_NT^{(0)}}{S}
   H_NT^{(0)}\mathbf{S}^{(1)} \rangle _{va}^{pq} = 
   \sum_{r} g_{pqra} \xi_{v}^{r} + \sum_{rs} g_{pqrs} \xi_{va}^{rs}
 - \sum_{rb} (g_{pbra} \xi_{vb}^{rq} \nonumber \\
 &+& g_{bqar} \xi_{vb}^{pr}) +  \sum_{br} \left[- t_b^p g_{qbar} 
 - t_b^q g_{pbra} \right] \xi_v^r 
 + \sum_{rs} g_{pqrs} t_a^s \xi_v^r + \sum_{bsr} \left[g_{bpsr} ( t_b^s \xi_{va}^{rq} 
 - t_b^r \xi_{va}^{sq} - t_a^s \xi_{bv}^{qr} - t_b^q \xi_{av}^{sr}) 
 + g_{bqsr} (t_a^r \xi_{bv}^{sp} \right. \nonumber \\
 &-& \left. t_b^p \xi_{va}^{sr}  - t_a^r \xi_{bv}^{ps} - t_a^s \xi_{bv}^{rp}
 + t_b^s \xi_{av}^{rp} - t_b^r \xi_{av}^{sp})\right]   
 + \sum_{rbc} g_{bcra} \left[t_c^r \xi_{bv}^{qp} - t_b^r \xi_{cv}^{qp} 
 + t_c^q \xi_{bv}^{pr} + t_b^q \xi_{cv}^{rp} + t_b^p \xi_{vc}^{rq} 
 - t_c^q \xi_{bv}^{rp} + t_{cb}^{qp} \xi_v^r \right] \nonumber \\ 
&+& g_{pbsr} \left [ (t_{ba}^{rq} - t_{ab}^{rq}) \xi_v^s 
 - t_{ab}^{qs} \xi_v^r \right] - g_{qbsr} t_{ab}^{sp}) \xi_v^r ]  
 + \sum_r g_{pr} t_{va}^{rq}
\label{HNbar-S1_2b-1}
\end{eqnarray}
\begin{eqnarray} 
  && \langle \contraction[0.4ex]{}{H}{_N}{T}\contraction[0.8ex]
     {}{H}{_NT^{(0)}}{T} \contraction[1.2ex] {}{H}{_NT^{(0)}T^{(0)}}{S}
     H_NT^{(0)}T^{(0)}\mathbf{S}^{(1)} \rangle _{va}^{pq} 
  = \sum_{brcs} g_{bcrs} \left[(t_{ab}^{qr} - t_{ab}^{rq} )\xi_{cv}^{sp} 
  + t_{ab}^{qs} \xi_{cv}^{rp} + t_{ab}^{rs} \xi_{cv}^{qp} 
  + t_{bc}^{qr} \xi_{av}^{sp} - t_{ab}^{qr} \xi_{cv}^{ps} + t_{ab}^{qp} (\xi_{cv}^{rs} 
  - \xi_{vc}^{rs}) \right. \nonumber \\ 
 &+& \left. t_{ab}^{rq} \xi_{vc}^{sp} + t_{ab}^{qs} \xi_{vc}^{rp} 
  - t_{bc}^{rq} \xi_{av}^{sp} - t_{ba}^{rs} \xi_{cv}^{qp}\right ] 
  + \sum_{rbs} \left[- g_{bprs} t_{a}^{r} t_{b}^{q} \xi_{v}^{s} 
  - g_{bqrs} t_{a}^{s} t_{b}^{p} \xi_{v}^{r} \right] 
  + \sum_{cbr} g_{cbar} t_c^q t_b^p \xi_{v}^{r} 
  + \sum_{brcs} g_{bcrs} [(-t_{ab}^{qr} t_c^p \nonumber \\
&+& t_{ab}^{qp} t_c^r )\xi_v^s + (- t_{ab}^{qp} t_c^s + t_{ab}^{qs} t_c^p) \xi_v^r 
 + (t_{ac}^{rp} t_b^q + t_{bc}^{qp} t_a^r + t_{ab}^{rq} t_c^p) \xi_v^s  
 + (t_b^q \xi_{av}^{rs} + t_a^r \xi_{bv}^{qs} - t_b^r \xi_{va}^{sq}) t_c^p 
 + t_b^p \xi_{va}^{sq} t_c^r + t_a^r (- t_b^q \xi_{cv}^{sp} \nonumber \\
&+& t_b^s \xi_{cv}^{qp}) + t_b^q (t_c^r \xi_{av}^{sp} + t_a^r \xi_{cv}^{ps}).
\label{HNbar-S1_2b-2}
\end{eqnarray}
\end{widetext}

\begin{figure*}
\includegraphics[width=16.5cm, angle=0]{HN_bar_S1.eps}
  \caption{Double PRCC diagrams contributing to the terms 
  $\contraction[0.4ex]{}{H}{_N}{S} H_N\mathbf{S}^{(1)} 
 + \contraction[0.4ex]{}{H}{_N}{T} \contraction[0.8ex]
  {}{H}{_NT^{(0)}}{S}H_NT^{(0)}\mathbf{S}^{(1)}$ (left panel) and 
  $\contraction[0.4ex]{}{H}{_N}{T}\contraction[0.8ex]{}{H}{_NT^{(0)}}{T} 
  \contraction[1.2ex] {}{H}{_NT^{(0)}T^{(0)}}{S}H_NT^{(0)}T^{(0)}\mathbf{S}^{(1)}$ 
 (right panel) of Eq. (\ref{peqn_4}).}
\label{diag-t5s2}
\end{figure*}


\subsubsection{Folded diagrams}

\begin{figure}
 \includegraphics[width=6.5cm, angle=0]{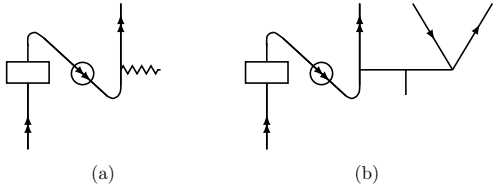}
	\caption{Folded diagrams contributing to PRCC Eqs. (\ref{peqn_3}) 
	(diagram (a)) and (\ref{peqn_4}) (diagram (b)).} 
 \label{diag-f}
\end{figure}

The terms on the right-hand sides of the PRCC Eqs.(\ref{peqn_3}) and 
(\ref{peqn_4}) are referred to as the renormalization terms in the CC 
equation of the one-valence systems. It is an important term, and its nonzero 
value distinguishes the PRCC equations of open-shell systems from 
the closed-shell systems. These contribute through the folded diagrams arising 
from the contraction of the energy with the CC operators. This contraction is
not possible in the case of closed-shell systems as the energy diagrams 
do not have free lines. Folded diagrams contributing to 
Eqs. (\ref{peqn_3}) and (\ref{peqn_4}) are given in Fig. \ref{diag-f} 
as diagrams (a) and (b), respectively.

\section{Polarizability for one-valence using PRCC}

In the PRCC, the electric dipole polarizability $\alpha$ of an atom or ion 
is defined as the expectation of the dipole operator with respect to the 
perturbed state $| \tilde{\Psi}_v \rangle$. For the one-valence atomic 
system
\begin{equation}
  \alpha = -\frac{\langle \tilde{\Psi}_v |\mathbf{D}|\tilde{\Psi}_v 
           \rangle}{\langle \Psi_v | \Psi_v \rangle}
  \label{alpha_1}
\end{equation}
Using the expression of $| \tilde{\Psi}_v \rangle$ from Eq. (\ref{ppsi_1}) and
retaining only the terms with first-order in $\lambda$, we get
\begin{eqnarray}
  \alpha &=& \frac{1}{\cal N}\langle \Phi_v |\bar{\mathbf{D}} 
             \left ( \mathbf{S}^{(1)} + \mathbf{T}^{(1)} + \mathbf{T}^{(1)}  
             S^{(0)} \right ) + \left( \mathbf{S}^{(1)} + \mathbf{T}^{(1)} 
             \right. \nonumber \\
         &&  + \left. \mathbf{T}^{(1)} S^{(0)} \right)^\dagger 
             \bar{\mathbf{D}} + {S^{(0)}}^\dagger \bar{\mathbf{D}} 
             \left ( \mathbf{S}^{(1)}  + \mathbf{T}^{(1)} 
             +  \mathbf{T}^{(1)} S^{(0)} \right ) 
                         \nonumber \\ 
         &&  + \left ( \mathbf{S}^{(1)} + \mathbf{T}^{(1)} 
             + \mathbf{T}^{(1)} S^{(0)} \right )^\dagger 
             \bar{\mathbf{D}} S^{(0)} | \Phi_v \rangle,
  \label{alpha_2}
\end{eqnarray}
where 
\begin{equation}
   {\cal N} = \langle \Phi_v|(e^{T^{(0)}}(1 + S^{(0)}))^\dagger
   (e^{T^{(0)}}(1 + S^{(0)}))| \Phi_v \rangle,
\end{equation}
is the normalization factor of the eigenstate $|\Psi_v\rangle$ and 
$\bar{\mathbf{D}}, = e^{{T^{(0)}}^\dagger} {\mathbf D} e^{T^{(0)}}$, is a 
dressed operator, which is a nonterminating series of the cluster
operator ${T^{(0)}}$. In the present work, however, we consider up to to
the second order term $\bar{\mathbf{D}} = \mathbf{D} + \mathbf{D} {T^{(0)}} 
+ {T^{(0)}}^\dagger \mathbf{D} + {T^{(0)}}^\dagger \mathbf{D} {T^{(0)}}$. 
The  higher order terms in $T^{(0)}$ have negligible contributions 
and this has been confirmed through detailed computations \cite{mani-10}.
In the CCSD approximation, Eq. (\ref{alpha_2}) can be written as
\begin{widetext}
\begin{eqnarray}
\alpha &= & \frac{1}{\cal N} \langle \Phi_v |
            \left ( {\mathbf D}\mathbf{S}_1^{(1)} +{\mathbf D} 
            \mathbf{S}_2^{(1)} + S_1^{(0)\dagger}{\mathbf D} \mathbf{S}_1^{(1)}
            + S_1^{(0)\dagger}{\mathbf D}\mathbf{S}_2^{(1)} 
            + S_2^{(0)\dagger}{\mathbf D} \mathbf{S}_1^{(1)}
            + S_2^{(0)\dagger}{\mathbf D} \mathbf{S}_2^{(1)} 
            +  S_1^{(0)\dagger}{\mathbf D} \mathbf{T}_1^{(1)} 
            + S_2^{(0)\dagger}{\mathbf D} \mathbf{T}_1^{(1)} \right. 
                     \nonumber  \\ 
       &&   \left. + S_2^{(0)\dagger}{\mathbf D} \mathbf{T}_2^{(1)} 
            + T_1^{(0)\dagger}{\mathbf D}\mathbf{S}_1^{(1)}
            + T_1^{(0)\dagger}{\mathbf D} \mathbf{S}_2^{(1)} 
            + T_2^{(0)\dagger} {\mathbf D}\mathbf{S}_2^{(1)}
            + {\mathbf D} \mathbf{T}^{(1)} 
            + T_1^{(0)\dagger} {\mathbf D}\mathbf{T}_1^{(1)}
            + T_1^{(0)\dagger} {\mathbf D} \mathbf{T}_2^{(1)}
            + T_2^{(0)\dagger} {\mathbf D} \mathbf{T}_1^{(1)} \right. 
                      \nonumber  \\
       && \left. + T_2^{(0)\dagger} {\mathbf D} \mathbf{T}_2^{(1)} \right)
            + {\text{H.c}} + {\mathbf D} \mathbf{T}_1^{(1)} S_1^{(0)} 
            + {\mathbf D} \mathbf{T}_1^{(0)}\mathbf{S}_1^{(1)} |\Phi_v \rangle.
  \label{alpha_3}
\end{eqnarray}
\end{widetext}
Here, the terms 
$S_1^{(0)\dagger} {\mathbf D} \mathbf{T}_2^{(1)} + \text{H.c.}$,
$T_2^{(0)\dagger} {\mathbf D} \mathbf{S}_2^{(1)} + \text{H.c.}$ and 
${\mathbf D}\mathbf{T}_2^{(1)} + \text{H.c.}$ are not included as these
do not contribute to the $\alpha$ of the one-valence system.


\subsection{Diagrams for $\alpha$}

\begin{figure}
 \includegraphics[width=7.0cm,height=12.0cm, angle=0]{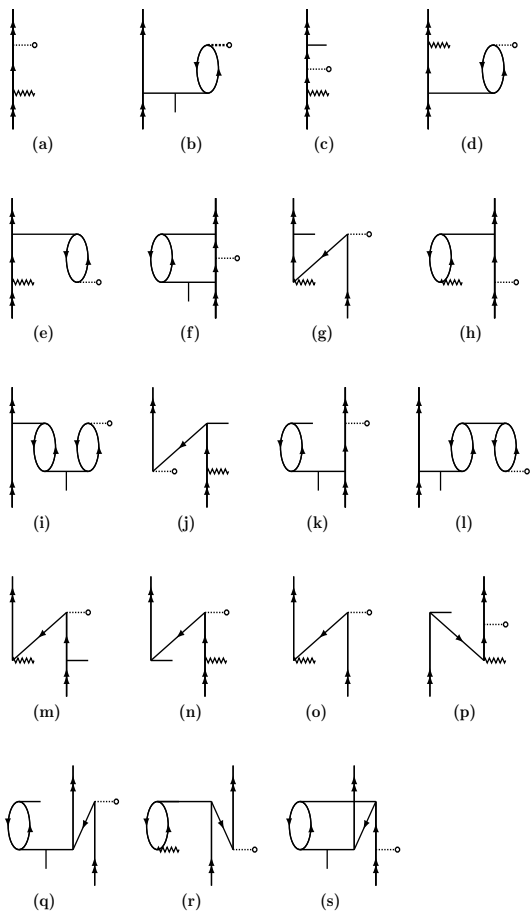}
 \caption{Some example polarizability diagrams for one-valence atomic system,
	  contributing to Eq. (\ref{alpha_3})}.
 \label{alpha-diag}
\end{figure}

There are 128 Goldstone diagrams which contribute to the Eq. (\ref{alpha_3}). 
And as example of the diagrams, in Fig. \ref{alpha-diag}, we show one 
diagram from each of the terms in the Eq. (\ref{alpha_3}). The diagrams from 
the Hermitian conjugate terms are, however, not shown as these are 
topologically equivalent. Among all the terms in the Eq. (\ref{alpha_3}), the 
first four terms, ${\mathbf D} \mathbf{S}_1^{(1)}$, 
${\mathbf D} \mathbf{S}_2^{(1)}$ and their hermitian conjugates, are expected 
to have the dominant contribution. The reason for this is the large 
magnitude of the one-valence cluster operators and the strong effect of 
the perturbation on these operators. More importantly, the terms
${\mathbf D}\mathbf{S}_1^{(1)} + \text{H.c}$. subsumes  the contributions from 
the Dirac-Fock (DF) and the random-phase-approximation (RPA). The diagrams 
of the ${\mathbf D}\mathbf{S}_1^{(1)}$ and ${\mathbf D}\mathbf{S}_2^{(1)}$ 
are shown in Fig. \ref{alpha-diag}(a) and (b), respectively.

Among the terms with two-orders of CC operators, 
$S_1^{(0)\dagger}{\mathbf D}\mathbf{S}_1^{(1)}$, 
$S_1^{(0)\dagger}{\mathbf D}\mathbf{S}_2^{(1)}$,
$S_2^{(0)\dagger} {\mathbf D}\mathbf{S}_1^{(1)}$, 
$S_2^{(0)\dagger}{\mathbf D} \mathbf{S}_2^{(1)}$ and
their H.c., are expected to give dominant contributions. The example diagrams 
of these four terms are shown in the Fig. \ref{alpha-diag}(c-f). The next 
important contributions are expected from the terms with 
one each of the  $T$ and $S$ operators;
$S_1^{(0)\dagger}{\mathbf D}\mathbf{T}_1^{(1)}$, 
$S_2^{(0)\dagger}{\mathbf D} \mathbf{T}_1^{(1)}$,
$S_2^{(0)\dagger}{\mathbf D}\mathbf{T}_2^{(1)}$, 
$T_1^{(0)\dagger}{\mathbf D} \mathbf{S}_1^{(1)}$,
$T_1^{(0)\dagger}{\mathbf D}\mathbf{S}_2^{(1)}$, 
$T_2^{(0)\dagger}{\mathbf D} \mathbf{S}_2^{(1)}$ and
their H.c., and ${\mathbf D}\mathbf{T}_1^{(1)}S_1^{(0)}$ 
and ${\mathbf D} T_1^{(0)} \mathbf{S}_1^{(1)}$. The representative diagrams 
from these terms are shown in Fig.\ref{alpha-diag}(g-n). The remaining terms,
${\mathbf D}\mathbf{T}_1^{(1)}$, 
$T_1^{(0)\dagger} {\mathbf D} \mathbf{T}_1^{(1)}$,
$T_1^{(0)\dagger} {\mathbf D}\mathbf{T}_2^{(1)}$, 
$T_2^{(0)\dagger} {\mathbf D} \mathbf{T}_1^{(1)}$,
$T_2^{(0)\dagger} {\mathbf D}\mathbf{T}_2^{(1)}$ and their H.c., having
two-orders of closed-shell operator, are expected to have the lowest 
contribution to $\alpha$. This is due to the small magnitudes of these 
operators for the open-shell systems. Some representative diagrams from 
these are shown in Fig.\ref{alpha-diag}(o-s).

\section{Basis set convergence}

 A basis set which provides a good description of the single-electron wave 
functions and energies is essential to obtain accurate and reliable results 
using the PRCC theory. In this work, we have used Gaussian-type orbitals 
(GTOs) \cite{mohanty-91} as the single-electron basis set. To ensure the 
accuracy of the basis set, the basis parameters are optimized to match the
orbital energies as well as the self-consistent-field energies with the 
GRASP2K \cite{jonsson-13} results. We achieved an excellent agreement between 
the GTO and GRASP2K energies. A detailed analysis and comparison of the 
energies of the group-13 elements is reported in our recent 
work \cite{ravi-20}. To improve the quality of the basis set further, we 
incorporate the effects of Breit interaction, vacuum polarization (VP) and the 
self-energy (SE) corrections in the basis set generation. For the Breit 
interaction, we employ the expression given in Ref. \cite{grant-80} and 
incorporate it at the level of orbital basis as well as the PRCC 
calculations. The effects of the vacuum polarization is considered using the 
Uehling potential \cite{uehling-35}, modified to incorporate the finite size 
effects of nucleus \cite{fullerton-76}. The self-energy corrections 
to the orbitals are incorporated through the model Lamb-shift operator 
introduced by Shabaev {\em et al.} \cite{shabaev-13}, and were calculated 
using the code QEDMOD \cite{shabaev-15}.



Owing to the mathematically incomplete nature of the GTO 
basis \cite{grant-06}, it is essential to check the convergence of 
properties with the basis size. In Table \ref{con-alpha}, we demonstrate the 
convergence of $\alpha$ for Al and In by listing $\alpha$ for both the 
fine structure states as a function of the basis size. For these
calculations we have used the Dirac-Coulomb (DC) Hamiltonian as using the
DCB Hamiltonian is computationally more expensive. As evident from the table, 
to achieve the convergence of $\alpha$, we start with a moderate 
basis size by considering up to the $h$ symmetry. And, then systematically 
increase the number of orbitals in each symmetry until the change in $\alpha$ 
is $\leqslant 10^{-3}$ a.u. For example, in the case of Al, the change in 
$\alpha$ is $2\times10^{-8}$ a.u. when the basis set is augmented from 175 
to 181 orbitals. So, we consider the basis set with the 175 orbitals as the 
optimal set and use it in further calculations to incorporate the effects 
of the Breit interaction and QED corrections. The same approach is also 
adopted to achieve the convergence of excitation energies, however, with a 
key difference. In this case, the basis set includes orbitals from
$j$-symmetry also. The convergence trends of the excitation energies 
and $\alpha$ are shown in the Fig. \ref{fig-ee} (a) and (b), respectively.
As discernible from the figure, both the excitation energy
and $\alpha$ converge well with the basis size.
\begin{table}
\caption{Convergence trend of $\alpha$ calculated using the 
	Dirac-Coulomb Hamiltonian as function of basis size.} 
  \label{con-alpha}
  \begin{ruledtabular}
  \begin{tabular}{cccc}
   Basis & Orbitals  &  $\alpha$ \\
   \cline{1-2}    \cline{3-4}

        & & $^2P_{1/2}$  & $^2P_{3/2}$ \\       
       \multicolumn{4}{c}{Al}\\         
       \hline  
  $98$  &  18s, 18p,   9d,  6f,  4g,  3h & $-58.852$ &  $-64.185$   \\         
  $120$ &  20s, 20p,  11d,  8f,  6g,  5h & $-59.433$ &  $-65.183$   \\       
  $142$ &  22s, 22p,  13d, 10f,  8g,  7h & $-58.762$ &  $-64.461$   \\  
  $164$ &  24s, 24p,  15d, 12f, 10g,  9h & $-58.347$ &  $-63.985$   \\         
  $175$ &  25s, 25p,  16d, 13f, 11g, 10h & $-58.273$ &  $-63.896$   \\      
  $181$ &  27s, 27p,  16d, 13f, 11g, 10h & $-58.273$ &  $-63.896$   \\                        

       \multicolumn{4}{c}{In}\\                 
       \hline 
  $110$  & 18s, 18p, 13d,  6f,  5g,  4h  & $-67.934$ & $-87.128$   \\
  $132$  & 20s, 20p, 15d,  8f,  7g,  6h  & $-68.339$ & $-87.489$   \\        
  $154$  & 22s, 22p, 17d, 10f,  9g,  8h  & $-67.203$ & $-86.093$   \\ 
  $176$  & 24s, 24p, 19d, 12f, 11g, 10h  & $-64.678$ & $-82.842$   \\  
  $187$  & 25s, 25p, 20d, 13f, 12g, 11h  & $-64.024$ & $-82.036$   \\ 
  $192$  & 26s, 26p, 21d, 13f, 12g, 11h  & $-64.027$ & $-81.996$   \\
  $197$  & 27s, 27p, 22d, 13f, 12g, 11h  & $-64.027$ & $-81.996$   \\ 
       
 \end{tabular}  
 \end{ruledtabular}
\end{table}            


\section{Results and Discussion}


\subsection{Excitation energies}

The excitation energy of a state $|\Psi_w\rangle$ is defined as
\begin{equation}
	\Delta E_{w} =  E_{w} - E_{v}, 
\end{equation}
where $E_{v}$ is the energy of the ground state wavefunction, and obtained 
from the solution of Eq. (\ref{psi_2}) for $3p_{1/2}$ and $5p_{1/2}$ states 
for Al and In, respectively. And, $E_{w}$ is the energy of an excited 
state $|\Psi_w\rangle$. In the RCC, $E_{w}$ is given by \cite{mani-10}
\begin{equation}
  E_w = \langle \Phi_w |{\bar H^{\rm DCB}}(1 + S^{(0)})| \Phi_w \rangle, 
\end{equation}
where $|\Phi_w \rangle$ is an excited Dirac-Fock state. In the 
Table \ref{tab-ee}, we have listed the energy of the ground state and
the excitation energies of a few low-lying states of Al and In. For 
comparison, the experimental values from NIST \cite{nist} are also listed in
the table. For Al, our theoretical results are in excellent agreement with 
the experimental data for all states. The maximum relative error is 0.26\%, 
in the case of $3p_{1/2}$ state. For In also we observe the same trend of 
relative errors except for the state $5p_{3/2}$, where the error is 9\%. 
This could be attributed to the correlation effects from higher energy 
configurations not included in the present work due to divergence issues.

\begin{figure}
\includegraphics[width=8.0cm, angle=-90]{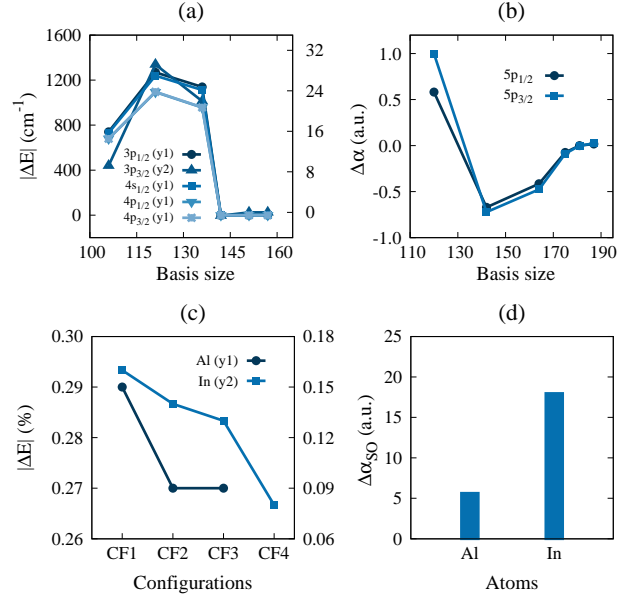}
\caption{Convergence of excitation energies (panel (a)) and dipole 
	polarizability (panel (b)) as function of basis size for Al. 
	Percentage change in the ground state energies of Al and In, 
	panel (c). Difference in the $\alpha$ values of spin-orbit 
	splitted states, $^2P_{1/2}$ and $^2P_{3/2}$, of Al and In.}
\label{fig-ee}
\end{figure}

To discern the electron correction effects as a function of configurations
included in the computations, energies are computed with different 
configuration spaces in steps.  For this we start with the ground state 
configuration in the configuration space and include the higher energy 
configurations in subsequent steps. For Al, we start with $3s^23p$ and 
refer to this as CF1. Then, we include two configurations $3s^24s$ and 
$3s^24p$ in two subsequent calculations (CF2 and CF3),
respectively. The inclusion of the configuration $3s^23d$, however, leads 
to the divergence in the FSRCC computations due to small energy denominator, 
and hence, we do not compute the excitation energy of $3d$. For In, 
$5s^25p$ (CF1) is the starting configuration and the 
excited state configurations $5s^26s$, $5s^26p$ and $5s^25d$ are 
included  in the later computations with configuration spaces identified as
CF2, CF3 and CF4, respectively. The trend of contributions from the higher 
energy configurations to the ground state energies of Al and In is shown 
in the Fig. \ref{fig-ee}(c). As we observe from the figure, for both the 
atoms, the relative error decreases with the inclusion of higher energy 
configurations. The reason for this is attributed to the better inclusion
of the {\em core-valence} and {\em valence-valence} correlations with 
larger configuration space.
\begin{table}
\caption{Energy (cm$^{-1}$) of the ground state and the excitation energies 
	of low-lying atomic states of Al and In.}
\label{tab-ee}
\begin{ruledtabular}
\begin{tabular}{lcc}
	States &  RCC results  & NIST\cite{nist} \\
  \hline  
	&  Al &   \\
  \hline	   
 $3s^2\ 3p_{1/2}$  & 48147.69  & 48275.20   \\
 $3s^2\ 3p_{3/2}$  &   111.93  &   112.06  \\
 $3s^2\ 4s_{1/2}$  & 25363.81  & 25347.76  \\
 $3s^2\ 4p_{1/2}$  & 32927.34  & 32949.81  \\
 $3s^2\ 4p_{3/2}$  & 32938.49  & 32965.64  \\
	   & In &   \\
	  \hline	
 $5s^2\ 5p_{1/2}$  & 46633.75  & 46670.20   \\
 $5s^2\ 5p_{3/2}$  &  2411.59  &  2212.59  \\
 $5s^2\ 6s_{1/2}$  & 24413.48  & 24372.96  \\
 $5s^2\ 6p_{1/2}$  & 31864.31  & 31816.96  \\
 $5s^2\ 6p_{3/2}$  & 32179.48  & 32115.22  \\ 
 $5s^2\ 5d_{3/2}$  & 32912.20  & 32892.21  \\
 $5s^2\ 5d_{5/2}$  & 32921.59  & 32915.54  \\  
\end{tabular}
\end{ruledtabular}
\end{table}

\subsection{Polarizability}

We list the value of $\alpha$ for ground state, $^2P_{1/2}$, and SO-coupled 
excited state, $^2P_{3/2}$, in the Tables \ref{pol_p1} and \ref{pol_p2}, 
respectively. For comparison we have also listed the other theoretical and 
experimental results from previous works. 
The results listed as PRCC are using the DC Hamiltonian and the converged 
bases with orbitals $25s25p16d13f11g10h$ and $26s26p21d13f12g11h$
for Al and In, respectively. And, the results listed as 
PRCC+Breit+QED incorporate the effects of Breit and QED corrections. The 
values listed as {\em estimated} refers to the value after incorporating 
the estimated contributions from the $i$, $j$ and $k$-symmetry orbitals. 
To quantify the effects of electron correlations from the nonlinear terms 
in the PRCC, the contributions from the DF and LPRCC are provided separately.

From the tables, we observe three important trends in the DF, LPRCC 
and PRCC $\alpha$ values for Al and In. First, except for the $^2P_{3/2}$ state 
of In, the LPRCC values are smaller than the DF values. This could 
be attributed to the contraction of the core with the inclusion of 
correlation effects within the LPRCC. 
Second, for both the atoms the PRCC values are larger than the DF. 
This is due to the contribution of electron correlations from the nonlinear terms. 
On close examination, we find that the nonlinear terms with one each of the
perturbed and unperturbed CC operators, viz, 
$\contraction[0.4ex]{}{H}{_N}{T}\contraction[0.7ex]{}{H}{_NT_0}
{\mathbf{T}_1^{(1)}} H_N\mathbf{T}^{(1)}S^{(0)}$  
and
$\contraction[0.4ex]{}{H}{_N}{T} \contraction[0.7ex] {}{H}{_NT_0}
{\mathbf{T}_1^{(1)}} H_N{T^{(0)}}\mathbf{S}^{(1)}$, 
contribute the most. And third, the difference between $\alpha$ of the
fine-structure states, $\alpha_{\text{FS}}$, of In is more than three times 
larger than Al. This is shown in the Fig. \ref{fig-ee}(d). The reason for this 
could be the larger difference in the radial extents of the
$^2P_{1/2}$ and $^2P_{3/2}$ states in In. In the DF computations,
$\langle r\rangle_{^2P_{3/2}} - \langle r\rangle_{^2P_{1/2}} = 0.138$ a.u. 
for In, however, it is only 0.007 a.u. for Al. 
\begin{figure}
\includegraphics[width=8.0cm, angle=-90]{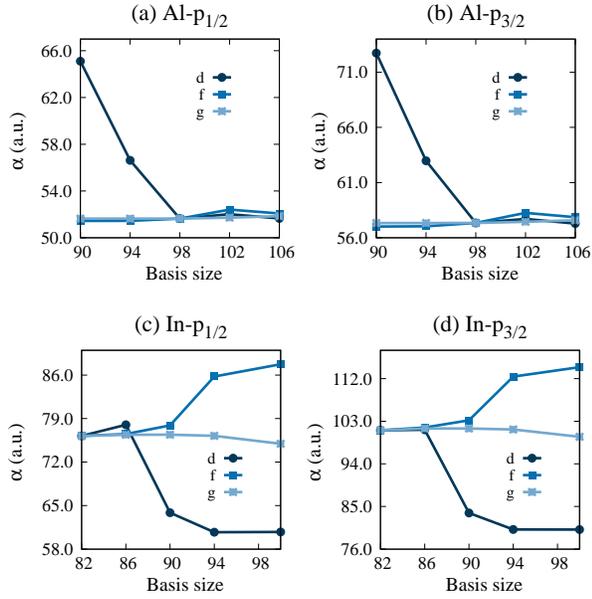}
\caption{Trend of contributions to $\alpha$ from virtual orbitals 
	for Al (panels (a) and (b)) and In (panels (c) and (d)) as
	basis is augmented.}
\label{fig-virt}
\end{figure}

\begin{table}
    \caption{The final value of $\alpha$ (a.u.) for $^2P_{1/2}$ from our 
     calculations compared with the other theory and experimental 
     results.}
  \label{pol_p1}
  \begin{center}
  \begin{ruledtabular}
  \begin{tabular}{cccr}
       Our results    & Method & Others &  Method \\
      \cline{1-4}
    \multicolumn{4}{c}{Al} \\
    \hline
              $57.083$   & DF       &  $55.4\pm2.2$\cite{fleig-05}      & MRCI     \\
              $51.537$   & LPRCC    &  $57.74$ \cite{lupinetti-05}      & CCSD(T)  \\
              $58.273$   & PRCC     &  $57.8\pm1.0$\cite{buchachenko-10}& SI-SOCI  \\
              $58.690$   & PRCC+Br. &  $58.0\pm0.4$\cite{fuentealba-04} & CCSD(T)  \\
	      $58.691$   & PRCC+Br.+QED & $61.0$\cite{chu-04}& SCI-DFT  \\
              $58.697$   & Est.        &  $46\pm2$\cite{milani-90}  & Exp.   \\
	      $58.70(59)$ & Reco.      & $55.3\pm5.5$\cite{sarkisov-06}   & Exp.  \\
    \multicolumn{4}{c}{In} \\
    \cline{1-4}
    \hline
              $62.756$    & DF        & $61.9\pm1.2$\cite{fleig-05}        & MRCI    \\
              $58.544$    & LPRCC     & $62.0\pm1.9 $\cite{borschevsky-12} & FSCC(T) \\
              $64.027$    & PRCC      & $61.5\pm5.6$\cite{safronova-13b}   & CCSD(T) \\
              $64.246$    & PRCC+Br.  & $66.4\pm5.0$\cite{buchachenko-10} & SI-SOCI  \\
              $64.269$    & PRCC+Br.+QED & $70.3$\cite{chu-04} & SCI-DFT    \\
              $64.228$    & Est.  & $68.7\pm8.1$\cite{guella-84}  & Exp.    \\
	      $64.23(64)$ & Reco. & $62.1\pm6.1$\cite{lei-15}   & Exp.    \\
  \end{tabular}
  \end{ruledtabular}
  \end{center}
\end{table}

\subsubsection{$^2P_{1/2}$}

For $^2P_{1/2}$ of Al, there are two experimental results of $\alpha$.  
However, there is a large difference between the reported values. The latest
experimental result of $\alpha$ given in the Ref. \cite{sarkisov-06} 
is $\approx 20$\% larger than the previous result reported in the 
Ref. \cite{milani-90}. In addition, there is significant difference in the 
experimental errors. The measurement in Ref. \cite{sarkisov-06} has an 
experimental error of $\approx 10$\%, whereas in the Ref. \cite{milani-90} it 
is $\approx 4$\%. Our recommended value, 58.70, is $\approx 6$\% larger than 
the Ref. \cite{sarkisov-06}. 
From the previous theoretical studies, there are five results for comparison. 
These include two coupled-cluster results, similar to method we have employed
in the present work. However, with a key difference in the calculation of 
$\alpha$: the two previous works used finite-field method. Like in the
experimental results, here as well, there is a variation in the reported 
values. There is a difference of about 10\% between the 
lowest \cite{fleig-05} and the highest \cite{chu-04} reported values. 
Although Refs. \cite{fleig-05} and  \cite{buchachenko-10} adopt the
same quantum many-body method, the value in Ref.  \cite{buchachenko-10} is 
larger than Ref. \cite{fleig-05}.
Our PRCC value, 58.27, is in good agreement with the CCSD(T) 
calculations, Refs. \cite{lupinetti-05} and \cite{fuentealba-04}, and 
SI-SOCI result \cite{buchachenko-10}. Our recommended value, 58.70, is
on the higher side of these results. The reason for this is attributed 
to the inclusion of the contributions from the Breit and QED corrections 
and the large basis sets in our calculations. The DF value

For In also there are two experimental results for ground state and, like Al, 
they differ by large amount -- the recent 
measurement by molecular-beam electric deflection technique \cite{lei-15} 
is about 10.6\% larger value than the Ref. \cite{guella-84}. 
Our recommended value, 64.23, lies between the two 
results. Among the previous calculations, in terms of methods adopted, the 
calculations by Borschevsky {\em et al.} \cite{borschevsky-12} and 
Safronova {\em et al.} \cite{safronova-13b} are closed to ours. Considering 
the error bars, our recommended value, 64.23, is in good agreement with these 
calculations. The reason for a small difference could be attributed 
to the basis set difference and the contributions from the Breit and QED 
corrections. The other two results are using the CI based 
calculations. The result, 66.4, from Ref.  \cite{buchachenko-10} is the 
largest among all the results and differ by about 8\% from the smallest 
value 61.5, Ref. \cite{fleig-05}.

\subsubsection{$^2P_{3/2}$}

Unlike the $^2P_{1/2}$ state, the static dipole polarizability for $^2P_{3/2}$ 
state will also have the contributions from the anisotropy components 
associated with magnetic quantum numbers $M_J = \pm 3/2$ and $\pm 1/2$. 
In Table \ref{pol_p2}, we have tabulated the average value, $\bar \alpha$,
of the polarizability. On close examination of the results, 
we observe three important differences in the trend of electron correlations 
in comparison to $^2P_{1/2}$ state. First, for both the atoms, the DF and 
LPRCC values are very close to each other. This indicates the less contraction 
of the core orbitals with the inclusion of the electron correlations. 
Second, the percentage contribution from the nonlinear terms in PRCC is 
less than $^2P_{1/2}$. And third, the overall Breit+QED correction has 
increased two-fold.

To the best of our knowledge, for both the atoms, there is no data on 
$\alpha$ from the experiments for $^2P_{3/2}$. And, from the previous 
calculations, there are few data from relativistic calculations which are 
listed in the table. In Refs. \cite{fleig-05} and \cite{borschevsky-12}, a
coupled-cluster method is employed to obtain the energy of Al and In, 
respectively,  and then $\alpha$ is calculated using the finite-field approach.
In Ref. \cite{buchachenko-10}, however, a configuration interaction
method is combined with finite-filed approach to calculate the $\alpha$ 
for In. For Al, our LPRCC result is within the error bars of the 
Refs. \cite{fleig-05, borschevsky-12}. Our recommended value, 64.69, is 
however larger than both the references. The reason for this is attributed 
to the large correlation effects from nonlinear terms in the PRCC theory. 
A sizable combined contribution from Breit+QED is also observed. The same 
trend of comparison with previous results is also observed for In. 
Here, however, there is a variation in the previous results 
and associated theoretical uncertainties. Our recommend value, 82.50, 
is within the error bar of the Ref. \cite{buchachenko-10}.
\begin{table}
  \caption{The value of $\bar{\alpha}$($p_{3/2}$) (a.u.) from our 
	calculations compared with the other theory and experimental results.}
  \label{pol_p2}
  \begin{center}
  \begin{ruledtabular}
  \begin{tabular}{cccr}
       Our results    & Method & Others &  Method \\
      \cline{1-4}
    \multicolumn{4}{c}{Al} \\
    \hline

               $57.655$ &    DF       & $55.9\pm2.2$\cite{fleig-05}  & MRCI \\
               $57.421$ &    LPRCC    & $58.0\pm1.0$\cite{buchachenko-10}  & SI-SOCI  \\
               $63.896$ &    PRCC     &           \\
               $64.703$ &    PRCC+Br.             \\
               $64.704$ &    PRCC+Br.+QED             \\
               $64.693$ &     Est.                \\
                   $64.69(65)$ &    Reco.             \\
    \multicolumn{4}{c}{In} \\
    \cline{1-4}
    \hline
               $76.157$ &  DF        &  $69.7\pm1.4$\cite{fleig-05}        & MRCI \\
               $76.586$ &  LPRCC     &  $74.4\pm8.0$\cite{buchachenko-10}  & SI-SOCI \\
               $81.996$ &  PRCC      &  $69.6\pm3.5$\cite{borschevsky-12}  &  FSCC(T)\\
               $82.545$ &  PRCC+Br.  &  \\
               $82.493$ &  PRCC+Br.+QED  &  \\
               $82.500$ &   Est. \\
               $82.50(83)$  & Reco. \\

  \end{tabular}
  \end{ruledtabular}
  \end{center}
\end{table}

\subsection{Electron correlations}

Next, we analyze and present the different electron correlations effects
incorporated in the calculations of $\alpha$. For this, we separate the
expression in Eq. (\ref{alpha_3}) into six different terms and give
their contributions in the Table \ref{pol_tw}. As evident from the table,
for both the atoms, the leading order (LO) term is
$\mathbf{S}^{(1)\dagger}\mathbf{D} + \text{H.c}$.
The contribution from the LO term is $\approx 146$\%(138\%) and
$\approx 141$\%(131\%) of the PRCC value for the $^2P_{1/2}$($^2P_{3/2}$)
state of Al and In, respectively. That is, the contribution from the LO term
exceeds the total value. This is expected as it incorporates the results
from the DF term and core-polarization (CP) effects. Except for the
$^2P_{3/2}$ of Al, the next leading order (NLO) term  is
$\mathbf{S}{^{(1) \dagger}}\mathbf{D}S^{(0)}$ + H.c. It contributes
$\approx  -10.6$\%($-6.5$\%) and $-13.6$\%($-10.8$\%) for the
$^2P_{1/2}$($^2P_{3/2}$) state of Al and In, respectively. For the next to
NLO contribution, the terms
$\mathbf{S}{^{(1) \dagger}}\mathbf{D}S^{(0)}$ + H.c. and
$\mathbf{T}^{(1)\dagger}\mathbf{D}$ +  H.c. give nearly equal contributions.
Like the NLO term, the contributions from these terms are opposite in phase
to the LO contribution, and hence, reduces the total value of $\alpha$. It
is to be mentioned here that the contributions from the core-electrons to
$\alpha$ are important. This is unlike the properties of one-valence systems
without an external perturbation like the electromagnetic transitions. This
is reflected in the contribution from the term
$\mathbf{T}^{(1)\dagger}\mathbf{D}$ + H.c. for both the atoms.
\begin{table}
    \caption{Contributions to $\alpha$ (a.u.) from different terms in PRCC
	     theory for Al and In.}
    \label{pol_tw}
    \begin{center}
    \begin{ruledtabular}
    \begin{tabular}{lrrrr}
        Terms + h. c.
        & \multicolumn{1}{r}{$\text{Al}$}
        & & \multicolumn{1}{r}{\text{In}} &  \\
        \cline{2-3}  \cline{4-5}
        & $^2P_{1/2}$  & $^2P_{3/2}$ & $^2P_{1/2}$  & $^2P_{3/2}$     \\
        \hline
        $\mathbf{S}^{(1)\dagger}\mathbf{D} $
        & $85.3337$ & $88.4131$  & $90.0421$ & $107.8206$ \\
        $\mathbf{S}{^{(1)\dagger}}\mathbf{D}S^{(0)} $
        & $-9.1048$  & $-5.7448$ &  $-12.2584$ & $-11.7265$    \\
        $T^{(0) \dagger} \mathbf{D} \mathbf{S}{^{(1)}} $
        & $-1.8365$ & $-2.2499$ &  $-0.7316$ & $-0.9715$     \\
        $S{^{(0)\dagger}} \mathbf{DT^{(1)}} $
        & $-7.2333$ & $-7.2820$ &  $-5.5342$ & $-5.5331$ \\
        $\mathbf{T}^{(1)\dagger}\mathbf{D} $
        & $-7.6424$  & $-7.9950 $&  $-7.4631$ & $-7.1608$  \\
        $T{^{(0)\dagger}} \mathbf{DT^{(1)}} $
        & $0.9567$ & $1.1739$ &  $1.5013$ & $1.5164$  \\
        Normalization & $-0.9636$  &  $-0.9744$ & $-0.9767$ & $-0.9768$\\
        Total    & $58.2732$  & $63.8955$ &  $64.0271$ &  $81.9959$\\
    \end{tabular}
    \end{ruledtabular}
    \end{center}
\end{table}

\begin{figure}
 \includegraphics[width=7.0cm, angle=0]{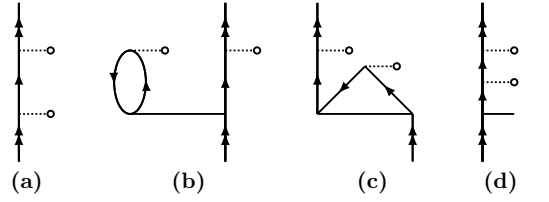}
 \caption{The DF (diagram (a)), CP (diagrams (b) and (c)) and VC
	(diagram (d)) terms subsumed in $D{\mathbf S}_1^{(1)}$.}
 \label{cp-pc}
\end{figure}

\begin{figure}
 \includegraphics[height=8.0cm, angle=-90]{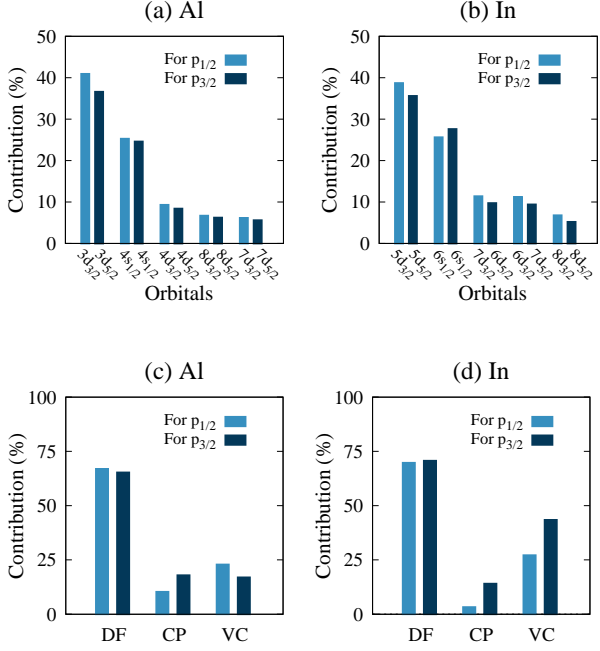}
	\caption{Five largest percentage contributions to
	$D{\mathbf S}_1^{(1)}$ + H.c. for $^2P_{1/2}$ and $^2P_{3/2}$
	states of Al (panel (a)) and In (panel (b)). The percentage
        contributions from DF, CP and VC for Al (panel (c)) and
        In (panel (d)).}
 \label{contri-ds11}
\end{figure}


\subsubsection{Dominant contributions}

To gain insights on the dominant contributions from the virtual orbitals,
we compute $\alpha$ using basis sets with selective addition of orbitals in
$d$, $f$ and $g$-symmetries. And, the results are plotted in the
Fig. \ref{fig-virt}. For Al the dominant contribution, as
discernible from the figure, is from the $d$-virtual orbitals.
A similar trend in Al$^+$ was reported in our previous work \cite{ravi-20}.
The same trend is also observed for In, however, with a key
difference. In this case the $f$-orbitals also contribute. And, this is
consistent with the trend reported in our previous work \cite{ravi-20} where
the $f$-virtual electrons were found to have dominant contribution due to
strong dipolar mixing with the core-electrons in the $4d$ orbital.

To quantify the orbital wise contributions, we identify the dominant
cluster amplitudes which  contribute to the LO term
$\mathbf{D}\mathbf{S}_1^{(1)}$ + H.c. As discernible from the
Fig. \ref{contri-ds11}, for Al, at 40.9\%(36.6\%) the cluster operators
with the virtual orbital $3d_{3/2}$($3d_{5/2}$) has the largest contribution
to the $^2P_{1/2}(^2P_{3/2})$ polarizability. This is due to the strong
dipolar mixing between the $3p$ and $3d$ orbitals. The second largest
contribution is observed from the cluster amplitudes with the $4s_{1/2}$
virtual orbital, the contribution is $\approx 25.3$\%(24.6\%) for the
$^2P_{1/2}(^2P_{3/2})$ state. The next three dominant contributions are from
the $4d$, $8d$ and $7d$ virtual orbitals, and together they contribute
$\approx 22.15$\% (20.18\%) for the $^2P_{1/2}(^2P_{3/2})$ state. A similar
trend is also observed in the case of In, where the first two dominant
contributions are from the $5d$ and $6s$-orbitals. They contribute
$\approx 38.7$\%(35.6\%) and 25.6\%(27.6\%), respectively,
for the $^2P_{1/2}(^2P_{3/2})$ state. In contrast to Al, the third and fourth
dominant contributions are of the same order and different $d$-electrons
contribute to $^2P_{1/2}$ and $^2P_{3/2}$ states.


\subsubsection{Core polarization, valence-virtual correlation and QED effects}

Next we assess the contributions from core polarization and pair correlation
effects to $\alpha$ of Al and In. The term $\mathbf{D}\mathbf{S}_1^{(1)}$
subsumes the contributions from DF, core-polarization (CP) and
valence-virtual correlation (VC) effects. The diagrams contributing to these
are shown in Fig. \ref{cp-pc}. The other dominant contribution to the
core-polarization is from the term $\mathbf{D}\mathbf{S}_2^{(1)}$, and
the corresponding diagram are shown in Fig. \ref{alpha-diag}(b) and its 
exchange. The contribution from the valence-virtual correlation is 
estimated by subtracting the contributions of diagrams (a), (b) and (c) in
Fig. \ref{cp-pc} from $\mathbf{D}\mathbf{S}^{(1)}$. The percentage
contributions of DF, CP and VC are shown in Fig. \ref{contri-ds11}.

For both the atoms, as to be expected, the DF has the largest contribution.
In terms of percentage, it constitutes $\approx 66.9$\%(65.2\%) and
69.7\%(70.6\%) of the $\mathbf{D}\mathbf{S}^{(1)}$ contribution for the
$^2P_{1/2}$($^2P_{3/2}$) state of Al and In, respectively. Between CP and VC,
except for $^2P_{3/2}$ state of Al, contribution from VC effect is larger
than CP and it is more significant in the case of In. In quantitative terms,
it constitutes $\approx 22.8$\%(16.9\%) and 27.1\%(43.4\%) of the
$\mathbf{D}\mathbf{S}^{(1)}$ for the $^2P_{1/2}$($^2P_{3/2})$ state of Al
and In, respectively. The CP contributions are $\approx$ 10.3\%(17.9\%)
and 3.2\%(13.9\%) of the $\mathbf{D}\mathbf{S}^{(1)}$ for
$^2P_{1/2}$($^2P_{3/2}$) state of Al and In.  It is, however, to be
emphasized that the CP contribution in In is smaller than Al. This indicates
a better screening of nuclear potential in In. The VC contribution, on the
contrary, is larger than Al.

\begin{figure}
 \includegraphics[height=8.0cm, angle=-90]{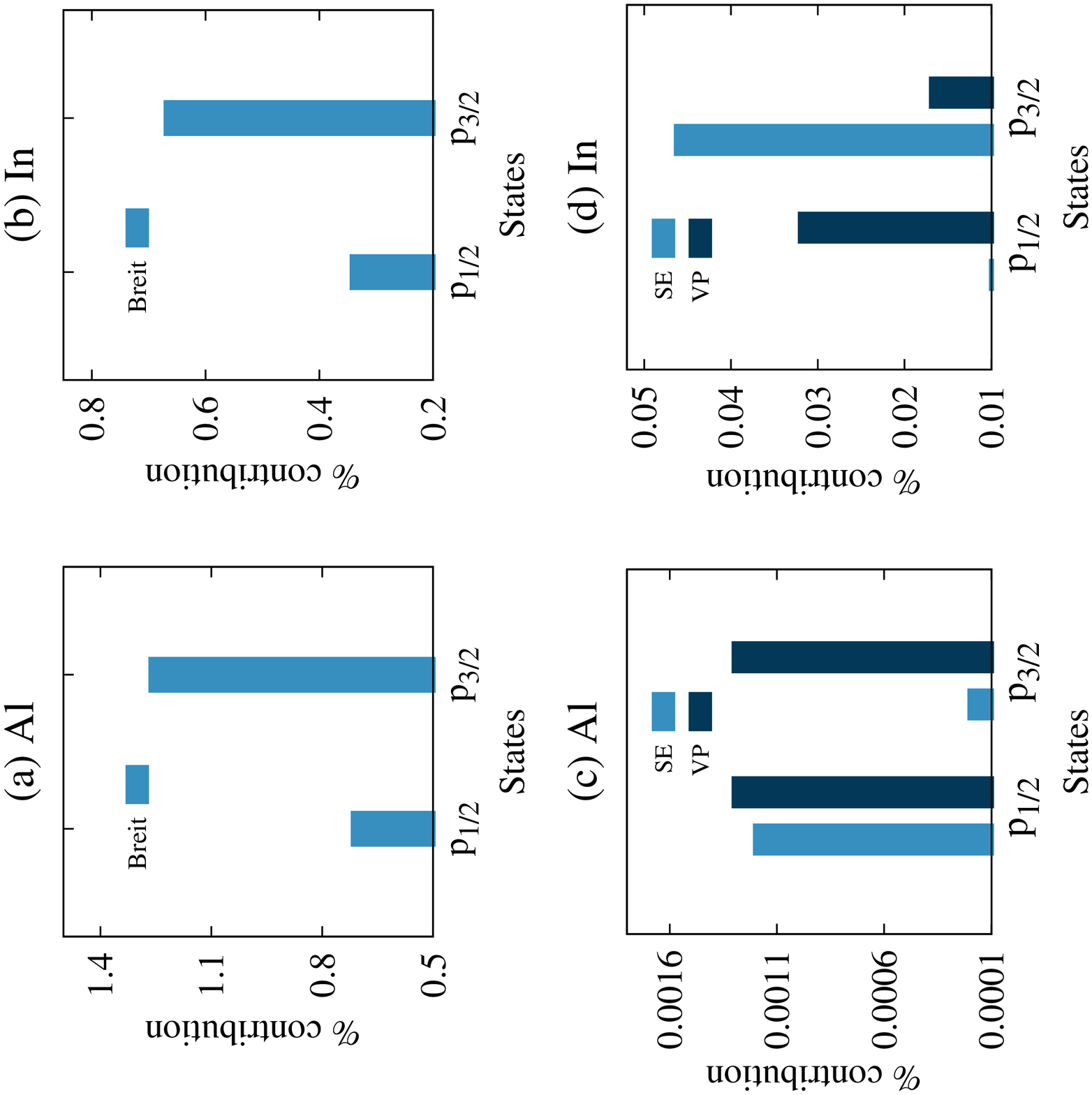}
 \caption{Contributions from the Breit interaction, vacuum polarization
     and the self-energy corrections for $^2P_{1/2}$ and $^2P_{3/2}$ states
     of Al and In.}
 \label{fig-qed}
\end{figure}

The contributions from the Breit interaction, vacuum polarization and the
self-energy corrections are listed in the Table \ref{pol_qed}. And, for
easy comparison, the contributions in percentage are plotted in the
Fig.\ref{fig-qed}.  As discernible from the figure, for the both states, the
Breit contribution in Al is larger than In. This is consistent with the
trend reported in our previous work \cite{ravi-20} where we found that, among
all the group-13 ions, Al$^+$ has the highest contribution. The largest
contribution is $\approx 1.3$\% for the $^2P_{3/2}$ state. For the VP and SE
contributions, in contrast to the trend of Breit contribution, these are
larger in In than Al. The largest contribution from VP is $\approx 0.3$\%
for the $^2P_{1/2}$ state, whereas, SE has the largest contribution of
$\approx 0.5$\%, for the case of $^2P_{3/2}$ state of In.
The largest combined contribution from  Breit interaction and QED
corrections is $\approx$ 1.3\%, in the case of $^2P_{3/2}$ state of Al.
Considering the need of accurate $\alpha$ from theory calculations, this is
a significant contribution and can not be ignored.

\begin{table}
    \caption{Contributions to $\alpha$ (a.u.) from Breit interaction,
	vacuum polarization and the self-energy corrections.}
    \label{pol_qed}
    \begin{center}
    \begin{ruledtabular}
    \begin{tabular}{lrrrr}
        & \multicolumn{1}{r}{$\text{Al}$}
        & & \multicolumn{1}{r}{\text{In}} &   \\
        \cline{2-3}  \cline{4-5}
        & $^2P_{1/2}$  & $^2P_{3/2}$  & $^2P_{1/2}$  & $^2P_{3/2}$     \\
        \hline
        DC          & $58.2732$ & $63.8955$  &  $64.0272$ & $81.9959$  \\
        Breit       & $ 0.4172$ & $ 0.8072$  &  $ 0.2192$ & $ 0.5490$  \\
        Self energy & $-0.0012$ & $-0.0002$  &  $ 0.0025$ & $-0.0380$  \\
        Vacuum pol. & $ 0.0013$ & $ 0.0013$  &  $ 0.0205$ & $-0.0139$  \\
    \end{tabular}
    \end{ruledtabular}
    \end{center}
\end{table}

\subsection{Theoretical uncertainty}

Based on the various approximations used in the computations of
$\alpha$, we have identified four sources of uncertainties. The first
source of uncertainty is associated with the basis set truncation. From the
convergence of $\alpha$ we observe that the change in
$\alpha$, with orbitals up to $h$-symmetry, is $\le 10^{-3}$ (a.u.) when the
optimal basis set is augmented. And, as listed in the Table \ref{pol_p2},
the largest overall contribution from the $i$, $j$ and $k$ symmetry orbitals
is in the case of $^2P_{3/2}$ of In and amounts to $\approx 0.06$\%.
Although the combined contribution from the orbitals with higher symmetries,
beyond the $k$-symmetry, is expected to be smaller we take 0.1\% as an
upper bound from this source. The second source of uncertainty
is the truncation of the dressed operator $\bar{\mathbf{D}}$
to ${\mathbf D}+{\mathbf D}T^{(0)}+{T^{(0)}}^\dagger {\mathbf D}
+ {T^{(0)}}^\dagger {\mathbf D}T^{(0)}$. To estimate the uncertainty from
this source, we use the findings from our previous work \cite{mani-10} where
we showed that the terms with third-order in $T^{(0)}$ and higher together
contribute less than 0.1\%. We take this as the upper bound from this source.
The third source of uncertainty is the truncation of CC operators to singles
and doubles. Among the higher excitations, triple excitations contribute the
most, and the dominant contribution is subsumed in the perturbative triples.
In our previous work \cite{ravi-20} on the dipole polarizability of group-13
ions, we had calculated the contributions from perturbative triples to be
$\approx$ 0.06\%  and 0.28\% for Al and In, respectively. Though the
cumulative contribution from the missing correlation effects in perturbative
triples and higher excitations is expected to be smaller, we take 0.56\%
as the upper bound from this source of uncertainty.
The last source of theoretical uncertainty is associated with the
frequency-dependent Breit interaction which is not included in our calculations.
To estimate an upper bound of this source we use the results in our previous
work \cite{chattopadhyay-14}, where using GRASP2K we estimated
an upper bound of $0.13$\% for Ra. As Al and In are lighter
than Ra, the contribution is expected to be smaller, we take
0.13\% as the uncertainty from this source. There could be other sources of
theoretical uncertainties, such as the higher order coupled perturbation of
vacuum polarization and self-energy terms, etc. These, however, have much
smaller contributions and their combined uncertainty could be below 0.1\%.
On combining the upper bounds of all four sources of uncertainties, we estimate
a theoretical uncertainty of 1\% in the recommended values of $\alpha$.

\section{Conclusion}

 We have developed a relativistic coupled-cluster theory based method to
compute the properties of one-valence atoms and ions with an external
perturbation. We employ this method to calculate the electric dipole
polarizability of ground state and SO-splitted excited state of Al and In.
In addition, to test the quality of the wavefunctions, we also calculated
the excitation energy of few low lying states. To improve the accuracy of
$\alpha$ further, contributions from the Breit interaction and QED
corrections are included. And, to ensure the convergence of $\alpha$ with
basis size, large bases up to $k$-symmetry are used.

 For the $^2P_{1/2}$ state, our recommended value lies within the
range of the previous theoretical results for both the atoms. In particular,
our results are closer to those reported in Refs. \cite{buchachenko-10} and
\cite{borschevsky-12} for Al and In, respectively. For the $^2P_{3/2}$ state,
however, our recommended value is lager than the previous values.
It is to be mentioned that our LPRCC values are closer to the previous results.
The reason for the larger PRCC values could be attributed to better inclusion
of correlation effects through the inclusion of nonlinear terms.

From the analysis of the electron correlations, we find that for both the
atoms, VC contribution is larger than CP. Between Al and In, the contribution
from CP decreases, however, VC effects are found to increase. In terms of
orbital contributions, for Al, the first two dominant contributions to $\alpha$
come from the $3p-3d$ and $3p-4s$ dipolar mixings. For In, however, they
are from the $5p-5d$ and $5p-6s$ mixings.  For the contribution
from the Breit interaction, the largest contribution is
$\approx$ 1.3\% of the DC value, observed in the case of $^2P_{3/2}$
state of Al. And, the largest contributions from the Uehling potential and
the self-energy corrections are, $\approx$ 0.3\% and 0.5\%, respectively,
in the case of $^2P_{1/2}$ and $^2P_{3/2}$ states of In.

\begin{acknowledgments}
We thank Siddhartha Chattopadhyay for useful suggestion on the manuscript. We 
would also like to thank Chandan, Suraj and Palki for useful discussions.
One of the authors, BKM, acknowledges the funding support from the
SERB (ECR/2016/001454). The results presented in the paper are based on the
computations using the High Performance Computing cluster, Padum, at the Indian
Institute of Technology Delhi, New Delhi.
\end{acknowledgments}

\bibliography{pol_1v}

\begin{thebibliography}{54}%
\makeatletter
\providecommand \@ifxundefined [1]{%
 \@ifx{#1\undefined}
}%
\providecommand \@ifnum [1]{%
 \ifnum #1\expandafter \@firstoftwo
 \else \expandafter \@secondoftwo
 \fi
}%
\providecommand \@ifx [1]{%
 \ifx #1\expandafter \@firstoftwo
 \else \expandafter \@secondoftwo
 \fi
}%
\providecommand \natexlab [1]{#1}%
\providecommand \enquote  [1]{``#1''}%
\providecommand \bibnamefont  [1]{#1}%
\providecommand \bibfnamefont [1]{#1}%
\providecommand \citenamefont [1]{#1}%
\providecommand \href@noop [0]{\@secondoftwo}%
\providecommand \href [0]{\begingroup \@sanitize@url \@href}%
\providecommand \@href[1]{\@@startlink{#1}\@@href}%
\providecommand \@@href[1]{\endgroup#1\@@endlink}%
\providecommand \@sanitize@url [0]{\catcode `\\12\catcode `\$12\catcode
  `\&12\catcode `\#12\catcode `\^12\catcode `\_12\catcode `\%12\relax}%
\providecommand \@@startlink[1]{}%
\providecommand \@@endlink[0]{}%
\providecommand \url  [0]{\begingroup\@sanitize@url \@url }%
\providecommand \@url [1]{\endgroup\@href {#1}{\urlprefix }}%
\providecommand \urlprefix  [0]{URL }%
\providecommand \Eprint [0]{\href }%
\providecommand \doibase [0]{http://dx.doi.org/}%
\providecommand \selectlanguage [0]{\@gobble}%
\providecommand \bibinfo  [0]{\@secondoftwo}%
\providecommand \bibfield  [0]{\@secondoftwo}%
\providecommand \translation [1]{[#1]}%
\providecommand \BibitemOpen [0]{}%
\providecommand \bibitemStop [0]{}%
\providecommand \bibitemNoStop [0]{.\EOS\space}%
\providecommand \EOS [0]{\spacefactor3000\relax}%
\providecommand \BibitemShut  [1]{\csname bibitem#1\endcsname}%
\let\auto@bib@innerbib\@empty
\bibitem [{\citenamefont {Dehmelt}(1982)}]{dehmelt_82}%
  \BibitemOpen
  \bibfield  {author} {\bibinfo {author} {\bibfnamefont {H.~G.}\ \bibnamefont
  {Dehmelt}},\ }\bibfield  {title} {\enquote {\bibinfo {title} {Mono-ion
  oscillator as potential ultimate laser frequency standard},}\ }\href
  {\doibase 10.1109/TIM.1982.6312526} {\bibfield  {journal} {\bibinfo
  {journal} {IEEE Trans. Instrum. Meas.}\ }\textbf {\bibinfo {volume}
  {IM-31}},\ \bibinfo {pages} {83} (\bibinfo {year} {1982})}\BibitemShut
  {NoStop}%
\bibitem [{\citenamefont {Chou}\ \emph {et~al.}(2010)\citenamefont {Chou},
  \citenamefont {Hume}, \citenamefont {Koelemeij}, \citenamefont {Wineland},\
  and\ \citenamefont {Rosenband}}]{chou-10}%
  \BibitemOpen
  \bibfield  {author} {\bibinfo {author} {\bibfnamefont {C.~W.}\ \bibnamefont
  {Chou}}, \bibinfo {author} {\bibfnamefont {D.~B.}\ \bibnamefont {Hume}},
  \bibinfo {author} {\bibfnamefont {J.~C.~J.}\ \bibnamefont {Koelemeij}},
  \bibinfo {author} {\bibfnamefont {D.~J.}\ \bibnamefont {Wineland}}, \ and\
  \bibinfo {author} {\bibfnamefont {T.}~\bibnamefont {Rosenband}},\ }\bibfield
  {title} {\enquote {\bibinfo {title} {Frequency comparison of two
  high-accuracy \mbox{${\mathrm{Al}}^ {+}$} optical clocks},}\ }\href {\doibase
  10.1103/PhysRevLett.104.070802} {\bibfield  {journal} {\bibinfo  {journal}
  {Phys. Rev. Lett.}\ }\textbf {\bibinfo {volume} {104}},\ \bibinfo {pages}
  {070802} (\bibinfo {year} {2010})}\BibitemShut {NoStop}%
\bibitem [{\citenamefont {Safronova}\ \emph {et~al.}(2011)\citenamefont
  {Safronova}, \citenamefont {Kozlov},\ and\ \citenamefont
  {Clark}}]{safronova-11b}%
  \BibitemOpen
  \bibfield  {author} {\bibinfo {author} {\bibfnamefont {M.~S.}\ \bibnamefont
  {Safronova}}, \bibinfo {author} {\bibfnamefont {M.~G.}\ \bibnamefont
  {Kozlov}}, \ and\ \bibinfo {author} {\bibfnamefont {Charles~W.}\ \bibnamefont
  {Clark}},\ }\bibfield  {title} {\enquote {\bibinfo {title} {Precision
  calculation of blackbody radiation shifts for optical frequency metrology},}\
  }\href {\doibase 10.1103/PhysRevLett.107.143006} {\bibfield  {journal}
  {\bibinfo  {journal} {Phys. Rev. Lett.}\ }\textbf {\bibinfo {volume} {107}},\
  \bibinfo {pages} {143006} (\bibinfo {year} {2011})}\BibitemShut {NoStop}%
\bibitem [{\citenamefont {Zuhrianda}\ \emph {et~al.}(2012)\citenamefont
  {Zuhrianda}, \citenamefont {Safronova},\ and\ \citenamefont
  {Kozlov}}]{zuhrianda-12}%
  \BibitemOpen
  \bibfield  {author} {\bibinfo {author} {\bibfnamefont {Z.}~\bibnamefont
  {Zuhrianda}}, \bibinfo {author} {\bibfnamefont {M.~S.}\ \bibnamefont
  {Safronova}}, \ and\ \bibinfo {author} {\bibfnamefont {M.~G.}\ \bibnamefont
  {Kozlov}},\ }\bibfield  {title} {\enquote {\bibinfo {title} {Anomalously
  small blackbody radiation shift in the \mbox{Tl${}^{+}$} frequency
  standard},}\ }\href {\doibase 10.1103/PhysRevA.85.022513} {\bibfield
  {journal} {\bibinfo  {journal} {Phys. Rev. A}\ }\textbf {\bibinfo {volume}
  {85}},\ \bibinfo {pages} {022513} (\bibinfo {year} {2012})}\BibitemShut
  {NoStop}%
\bibitem [{\citenamefont {Chen}\ \emph {et~al.}(2017)\citenamefont {Chen},
  \citenamefont {Brewer}, \citenamefont {Chou}, \citenamefont {Wineland},
  \citenamefont {Leibrandt},\ and\ \citenamefont {Hume}}]{chen-17}%
  \BibitemOpen
  \bibfield  {author} {\bibinfo {author} {\bibfnamefont {J.-S.}\ \bibnamefont
  {Chen}}, \bibinfo {author} {\bibfnamefont {S.~M.}\ \bibnamefont {Brewer}},
  \bibinfo {author} {\bibfnamefont {C.~W.}\ \bibnamefont {Chou}}, \bibinfo
  {author} {\bibfnamefont {D.~J.}\ \bibnamefont {Wineland}}, \bibinfo {author}
  {\bibfnamefont {D.~R.}\ \bibnamefont {Leibrandt}}, \ and\ \bibinfo {author}
  {\bibfnamefont {D.~B.}\ \bibnamefont {Hume}},\ }\bibfield  {title} {\enquote
  {\bibinfo {title} {Sympathetic ground state cooling and time-dilation shifts
  in an ${^{27}\mathrm{Al}}^{+}$ optical clock},}\ }\href {\doibase
  10.1103/PhysRevLett.118.053002} {\bibfield  {journal} {\bibinfo  {journal}
  {Phys. Rev. Lett.}\ }\textbf {\bibinfo {volume} {118}},\ \bibinfo {pages}
  {053002} (\bibinfo {year} {2017})}\BibitemShut {NoStop}%
\bibitem [{\citenamefont {Brewer}\ \emph
  {et~al.}(2019{\natexlab{a}})\citenamefont {Brewer}, \citenamefont {Chen},
  \citenamefont {Hankin}, \citenamefont {Clements}, \citenamefont {Chou},
  \citenamefont {Wineland}, \citenamefont {Hume},\ and\ \citenamefont
  {Leibrandt}}]{brewer_19a}%
  \BibitemOpen
  \bibfield  {author} {\bibinfo {author} {\bibfnamefont {S.~M.}\ \bibnamefont
  {Brewer}}, \bibinfo {author} {\bibfnamefont {J.-S.}\ \bibnamefont {Chen}},
  \bibinfo {author} {\bibfnamefont {A.~M.}\ \bibnamefont {Hankin}}, \bibinfo
  {author} {\bibfnamefont {E.~R.}\ \bibnamefont {Clements}}, \bibinfo {author}
  {\bibfnamefont {C.~W.}\ \bibnamefont {Chou}}, \bibinfo {author}
  {\bibfnamefont {D.~J.}\ \bibnamefont {Wineland}}, \bibinfo {author}
  {\bibfnamefont {D.~B.}\ \bibnamefont {Hume}}, \ and\ \bibinfo {author}
  {\bibfnamefont {D.~R.}\ \bibnamefont {Leibrandt}},\ }\bibfield  {title}
  {\enquote {\bibinfo {title} {$^{27}\mathrm{Al}^{+}$ quantum-logic clock with
  a systematic uncertainty below ${10}^{\ensuremath{-}18}$},}\ }\href {\doibase
  10.1103/PhysRevLett.123.033201} {\bibfield  {journal} {\bibinfo  {journal}
  {Phys. Rev. Lett.}\ }\textbf {\bibinfo {volume} {123}},\ \bibinfo {pages}
  {033201} (\bibinfo {year} {2019}{\natexlab{a}})}\BibitemShut {NoStop}%
\bibitem [{\citenamefont {Brewer}\ \emph
  {et~al.}(2019{\natexlab{b}})\citenamefont {Brewer}, \citenamefont {Chen},
  \citenamefont {Beloy}, \citenamefont {Hankin}, \citenamefont {Clements},
  \citenamefont {Chou}, \citenamefont {McGrew}, \citenamefont {Zhang},
  \citenamefont {Fasano}, \citenamefont {Nicolodi}, \citenamefont {Leopardi},
  \citenamefont {Fortier}, \citenamefont {Diddams}, \citenamefont {Ludlow},
  \citenamefont {Wineland}, \citenamefont {Leibrandt},\ and\ \citenamefont
  {Hume}}]{brewer_19b}%
  \BibitemOpen
  \bibfield  {author} {\bibinfo {author} {\bibfnamefont {S.~M.}\ \bibnamefont
  {Brewer}}, \bibinfo {author} {\bibfnamefont {J.-S.}\ \bibnamefont {Chen}},
  \bibinfo {author} {\bibfnamefont {K.}~\bibnamefont {Beloy}}, \bibinfo
  {author} {\bibfnamefont {A.~M.}\ \bibnamefont {Hankin}}, \bibinfo {author}
  {\bibfnamefont {E.~R.}\ \bibnamefont {Clements}}, \bibinfo {author}
  {\bibfnamefont {C.~W.}\ \bibnamefont {Chou}}, \bibinfo {author}
  {\bibfnamefont {W.~F.}\ \bibnamefont {McGrew}}, \bibinfo {author}
  {\bibfnamefont {X.}~\bibnamefont {Zhang}}, \bibinfo {author} {\bibfnamefont
  {R.~J.}\ \bibnamefont {Fasano}}, \bibinfo {author} {\bibfnamefont
  {D.}~\bibnamefont {Nicolodi}}, \bibinfo {author} {\bibfnamefont
  {H.}~\bibnamefont {Leopardi}}, \bibinfo {author} {\bibfnamefont {T.~M.}\
  \bibnamefont {Fortier}}, \bibinfo {author} {\bibfnamefont {S.~A.}\
  \bibnamefont {Diddams}}, \bibinfo {author} {\bibfnamefont {A.~D.}\
  \bibnamefont {Ludlow}}, \bibinfo {author} {\bibfnamefont {D.~J.}\
  \bibnamefont {Wineland}}, \bibinfo {author} {\bibfnamefont {D.~R.}\
  \bibnamefont {Leibrandt}}, \ and\ \bibinfo {author} {\bibfnamefont {D.~B.}\
  \bibnamefont {Hume}},\ }\bibfield  {title} {\enquote {\bibinfo {title}
  {Measurements of $^{27}\mathrm{Al}^{+}$ and $^{25}\mathrm{Mg}^{+}$ magnetic
  constants for improved ion-clock accuracy},}\ }\href {\doibase
  10.1103/PhysRevA.100.013409} {\bibfield  {journal} {\bibinfo  {journal}
  {Phys. Rev. A}\ }\textbf {\bibinfo {volume} {100}},\ \bibinfo {pages}
  {013409} (\bibinfo {year} {2019}{\natexlab{b}})}\BibitemShut {NoStop}%
\bibitem [{\citenamefont {van Wijngaarden}(1999)}]{wijngaarden-99}%
  \BibitemOpen
  \bibfield  {author} {\bibinfo {author} {\bibfnamefont {William~Arie}\
  \bibnamefont {van Wijngaarden}},\ }\bibfield  {title} {\enquote {\bibinfo
  {title} {Precision measurements of atomic polarizabilities},}\ }\href
  {\doibase 10.1063/1.59362} {\bibfield  {journal} {\bibinfo  {journal} {AIP
  Conference Proceedings}\ }\textbf {\bibinfo {volume} {477}},\ \bibinfo
  {pages} {305--321} (\bibinfo {year} {1999})},\ \Eprint
  {http://arxiv.org/abs/https://aip.scitation.org/doi/pdf/10.1063/1.59362}
  {https://aip.scitation.org/doi/pdf/10.1063/1.59362} \BibitemShut {NoStop}%
\bibitem [{\citenamefont {Khriplovich}(1991)}]{khriplovich-91}%
  \BibitemOpen
  \bibfield  {author} {\bibinfo {author} {\bibfnamefont {I.B.}\ \bibnamefont
  {Khriplovich}},\ }\href@noop {} {\emph {\bibinfo {title} {Parity
  Nonconservation in Atomic Phenomena}}}\ (\bibinfo  {publisher} {Gordon and
  Breach Science Publishers},\ \bibinfo {address} {Philadelphia},\ \bibinfo
  {year} {1991})\BibitemShut {NoStop}%
\bibitem [{\citenamefont {Griffith}\ \emph {et~al.}(2009)\citenamefont
  {Griffith}, \citenamefont {Swallows}, \citenamefont {Loftus}, \citenamefont
  {Romalis}, \citenamefont {Heckel},\ and\ \citenamefont
  {Fortson}}]{griffith-09}%
  \BibitemOpen
  \bibfield  {author} {\bibinfo {author} {\bibfnamefont {W.~C.}\ \bibnamefont
  {Griffith}}, \bibinfo {author} {\bibfnamefont {M.~D.}\ \bibnamefont
  {Swallows}}, \bibinfo {author} {\bibfnamefont {T.~H.}\ \bibnamefont
  {Loftus}}, \bibinfo {author} {\bibfnamefont {M.~V.}\ \bibnamefont {Romalis}},
  \bibinfo {author} {\bibfnamefont {B.~R.}\ \bibnamefont {Heckel}}, \ and\
  \bibinfo {author} {\bibfnamefont {E.~N.}\ \bibnamefont {Fortson}},\
  }\bibfield  {title} {\enquote {\bibinfo {title} {Improved limit on the
  permanent electric dipole moment of \mbox{Hg$^{199}$}},}\ }\href {\doibase
  10.1103/PhysRevLett.102.101601} {\bibfield  {journal} {\bibinfo  {journal}
  {Phys. Rev. Lett.}\ }\textbf {\bibinfo {volume} {102}},\ \bibinfo {pages}
  {101601} (\bibinfo {year} {2009})}\BibitemShut {NoStop}%
\bibitem [{\citenamefont {Anderson}\ \emph {et~al.}(1995)\citenamefont
  {Anderson}, \citenamefont {Ensher}, \citenamefont {Matthews}, \citenamefont
  {Wieman},\ and\ \citenamefont {Cornell}}]{anderson-95}%
  \BibitemOpen
  \bibfield  {author} {\bibinfo {author} {\bibfnamefont {M.~H.}\ \bibnamefont
  {Anderson}}, \bibinfo {author} {\bibfnamefont {J.~R.}\ \bibnamefont
  {Ensher}}, \bibinfo {author} {\bibfnamefont {M.~R.}\ \bibnamefont
  {Matthews}}, \bibinfo {author} {\bibfnamefont {C.~E.}\ \bibnamefont
  {Wieman}}, \ and\ \bibinfo {author} {\bibfnamefont {E.~A.}\ \bibnamefont
  {Cornell}},\ }\bibfield  {title} {\enquote {\bibinfo {title} {Observation of
  \mbox{B}ose-\mbox{E}instein condensation in a dilute atomic vapor},}\ }\href
  {\doibase 10.1126/science.269.5221.198} {\bibfield  {journal} {\bibinfo
  {journal} {Science}\ }\textbf {\bibinfo {volume} {269}},\ \bibinfo {pages}
  {198--201} (\bibinfo {year} {1995})}\BibitemShut {NoStop}%
\bibitem [{\citenamefont {Bradley}\ \emph {et~al.}(1995)\citenamefont
  {Bradley}, \citenamefont {Sackett}, \citenamefont {Tollett},\ and\
  \citenamefont {Hulet}}]{bradley-95}%
  \BibitemOpen
  \bibfield  {author} {\bibinfo {author} {\bibfnamefont {C.~C.}\ \bibnamefont
  {Bradley}}, \bibinfo {author} {\bibfnamefont {C.~A.}\ \bibnamefont
  {Sackett}}, \bibinfo {author} {\bibfnamefont {J.~J.}\ \bibnamefont
  {Tollett}}, \ and\ \bibinfo {author} {\bibfnamefont {R.~G.}\ \bibnamefont
  {Hulet}},\ }\bibfield  {title} {\enquote {\bibinfo {title} {Evidence of
  \mbox{B}ose-\mbox{E}instein condensation in an atomic gas with attractive
  interactions},}\ }\href {\doibase 10.1103/PhysRevLett.75.1687} {\bibfield
  {journal} {\bibinfo  {journal} {Phys. Rev. Lett.}\ }\textbf {\bibinfo
  {volume} {75}},\ \bibinfo {pages} {1687--1690} (\bibinfo {year}
  {1995})}\BibitemShut {NoStop}%
\bibitem [{\citenamefont {Davis}\ \emph {et~al.}(1995)\citenamefont {Davis},
  \citenamefont {Mewes}, \citenamefont {Andrews}, \citenamefont {van Druten},
  \citenamefont {Durfee}, \citenamefont {Kurn},\ and\ \citenamefont
  {Ketterle}}]{davis-95}%
  \BibitemOpen
  \bibfield  {author} {\bibinfo {author} {\bibfnamefont {K.~B.}\ \bibnamefont
  {Davis}}, \bibinfo {author} {\bibfnamefont {M.~O.}\ \bibnamefont {Mewes}},
  \bibinfo {author} {\bibfnamefont {M.~R.}\ \bibnamefont {Andrews}}, \bibinfo
  {author} {\bibfnamefont {N.~J.}\ \bibnamefont {van Druten}}, \bibinfo
  {author} {\bibfnamefont {D.~S.}\ \bibnamefont {Durfee}}, \bibinfo {author}
  {\bibfnamefont {D.~M.}\ \bibnamefont {Kurn}}, \ and\ \bibinfo {author}
  {\bibfnamefont {W.}~\bibnamefont {Ketterle}},\ }\bibfield  {title} {\enquote
  {\bibinfo {title} {\mbox{B}ose-\mbox{E}instein condensation in a gas of
  sodium atoms},}\ }\href {\doibase 10.1103/PhysRevLett.75.3969} {\bibfield
  {journal} {\bibinfo  {journal} {Phys. Rev. Lett.}\ }\textbf {\bibinfo
  {volume} {75}},\ \bibinfo {pages} {3969--3973} (\bibinfo {year}
  {1995})}\BibitemShut {NoStop}%
\bibitem [{\citenamefont {Lewenstein}\ \emph {et~al.}(1994)\citenamefont
  {Lewenstein}, \citenamefont {Balcou}, \citenamefont {Ivanov}, \citenamefont
  {L'Huillier},\ and\ \citenamefont {Corkum}}]{lewenstein-94}%
  \BibitemOpen
  \bibfield  {author} {\bibinfo {author} {\bibfnamefont {M.}~\bibnamefont
  {Lewenstein}}, \bibinfo {author} {\bibfnamefont {Ph.}\ \bibnamefont
  {Balcou}}, \bibinfo {author} {\bibfnamefont {M.~Yu.}\ \bibnamefont {Ivanov}},
  \bibinfo {author} {\bibfnamefont {Anne}\ \bibnamefont {L'Huillier}}, \ and\
  \bibinfo {author} {\bibfnamefont {P.~B.}\ \bibnamefont {Corkum}},\ }\bibfield
   {title} {\enquote {\bibinfo {title} {Theory of high-harmonic generation by
  low-frequency laser fields},}\ }\href {\doibase 10.1103/PhysRevA.49.2117}
  {\bibfield  {journal} {\bibinfo  {journal} {Phys. Rev. A}\ }\textbf {\bibinfo
  {volume} {49}},\ \bibinfo {pages} {2117--2132} (\bibinfo {year}
  {1994})}\BibitemShut {NoStop}%
\bibitem [{\citenamefont {Lewenstein}\ \emph {et~al.}(1995)\citenamefont
  {Lewenstein}, \citenamefont {Sali\`eres},\ and\ \citenamefont
  {L'Huillier}}]{lewenstein-95}%
  \BibitemOpen
  \bibfield  {author} {\bibinfo {author} {\bibfnamefont {Maciej}\ \bibnamefont
  {Lewenstein}}, \bibinfo {author} {\bibfnamefont {Pascal}\ \bibnamefont
  {Sali\`eres}}, \ and\ \bibinfo {author} {\bibfnamefont {Anne}\ \bibnamefont
  {L'Huillier}},\ }\bibfield  {title} {\enquote {\bibinfo {title} {Phase of the
  atomic polarization in high-order harmonic generation},}\ }\href {\doibase
  10.1103/PhysRevA.52.4747} {\bibfield  {journal} {\bibinfo  {journal} {Phys.
  Rev. A}\ }\textbf {\bibinfo {volume} {52}},\ \bibinfo {pages} {4747--4754}
  (\bibinfo {year} {1995})}\BibitemShut {NoStop}%
\bibitem [{\citenamefont {Blaga}\ \emph {et~al.}(2009)\citenamefont {Blaga},
  \citenamefont {Catoire}, \citenamefont {Colosimo}, \citenamefont {Paulus},
  \citenamefont {Muller}, \citenamefont {Agostini},\ and\ \citenamefont
  {DiMauro}}]{blaga-09}%
  \BibitemOpen
  \bibfield  {author} {\bibinfo {author} {\bibfnamefont {C.~I.}\ \bibnamefont
  {Blaga}}, \bibinfo {author} {\bibfnamefont {F.}~\bibnamefont {Catoire}},
  \bibinfo {author} {\bibfnamefont {P.}~\bibnamefont {Colosimo}}, \bibinfo
  {author} {\bibfnamefont {G.~G.}\ \bibnamefont {Paulus}}, \bibinfo {author}
  {\bibfnamefont {H.~G.}\ \bibnamefont {Muller}}, \bibinfo {author}
  {\bibfnamefont {P.}~\bibnamefont {Agostini}}, \ and\ \bibinfo {author}
  {\bibfnamefont {L.~F.}\ \bibnamefont {DiMauro}},\ }\bibfield  {title}
  {\enquote {\bibinfo {title} {Strong-field photoionization revisited},}\
  }\href {https://doi.org/10.1038/nphys1228} {\bibfield  {journal} {\bibinfo
  {journal} {Nature Phys.}\ }\textbf {\bibinfo {volume} {5}},\ \bibinfo {pages}
  {335} (\bibinfo {year} {2009})}\BibitemShut {NoStop}%
\bibitem [{\citenamefont {Lai}\ \emph {et~al.}(2018)\citenamefont {Lai},
  \citenamefont {Blaga}, \citenamefont {Xu}, \citenamefont {Fuest},
  \citenamefont {Rupp}, \citenamefont {Kling}, \citenamefont {Agostini},\ and\
  \citenamefont {DiMauro}}]{lai-18}%
  \BibitemOpen
  \bibfield  {author} {\bibinfo {author} {\bibfnamefont {Yu~Hang}\ \bibnamefont
  {Lai}}, \bibinfo {author} {\bibfnamefont {Cosmin~I.}\ \bibnamefont {Blaga}},
  \bibinfo {author} {\bibfnamefont {Junliang}\ \bibnamefont {Xu}}, \bibinfo
  {author} {\bibfnamefont {Harald}\ \bibnamefont {Fuest}}, \bibinfo {author}
  {\bibfnamefont {Philipp}\ \bibnamefont {Rupp}}, \bibinfo {author}
  {\bibfnamefont {Matthias~F.}\ \bibnamefont {Kling}}, \bibinfo {author}
  {\bibfnamefont {Pierre}\ \bibnamefont {Agostini}}, \ and\ \bibinfo {author}
  {\bibfnamefont {Louis~F.}\ \bibnamefont {DiMauro}},\ }\bibfield  {title}
  {\enquote {\bibinfo {title} {Polarizability effect in strong-field
  ionization: Quenching of the low-energy structure in
  \mbox{${\mathrm{C}}_{60}$}},}\ }\href {\doibase 10.1103/PhysRevA.98.063427}
  {\bibfield  {journal} {\bibinfo  {journal} {Phys. Rev. A}\ }\textbf {\bibinfo
  {volume} {98}},\ \bibinfo {pages} {063427} (\bibinfo {year}
  {2018})}\BibitemShut {NoStop}%
\bibitem [{\citenamefont {Karshenboim}\ and\ \citenamefont
  {Peik}(2010)}]{karshenboim-10}%
  \BibitemOpen
  \bibfield  {author} {\bibinfo {author} {\bibfnamefont {S.~G.}\ \bibnamefont
  {Karshenboim}}\ and\ \bibinfo {author} {\bibfnamefont {E}~\bibnamefont
  {Peik}},\ }\href@noop {} {\emph {\bibinfo {title} {Astrophysics, Clocks and
  Fundamental Constants, Lecture Notes in Physics}}}\ (\bibinfo  {publisher}
  {Springer},\ \bibinfo {address} {New York},\ \bibinfo {year}
  {2010})\BibitemShut {NoStop}%
\bibitem [{\citenamefont {Murphy}\ \emph {et~al.}(2007)\citenamefont {Murphy},
  \citenamefont {Webb},\ and\ \citenamefont {Flambaum}}]{murphy-07}%
  \BibitemOpen
  \bibfield  {author} {\bibinfo {author} {\bibfnamefont {M.~T.}\ \bibnamefont
  {Murphy}}, \bibinfo {author} {\bibfnamefont {J.~K.}\ \bibnamefont {Webb}}, \
  and\ \bibinfo {author} {\bibfnamefont {V.~V.}\ \bibnamefont {Flambaum}},\
  }\bibfield  {title} {\enquote {\bibinfo {title} {Comment on ``limits on the
  time variation of the electromagnetic fine-structure constant in the low
  energy limit from absorption lines in the spectra of distant quasars''},}\
  }\href {\doibase 10.1103/PhysRevLett.99.239001} {\bibfield  {journal}
  {\bibinfo  {journal} {Phys. Rev. Lett.}\ }\textbf {\bibinfo {volume} {99}},\
  \bibinfo {pages} {239001} (\bibinfo {year} {2007})}\BibitemShut {NoStop}%
\bibitem [{\citenamefont {Fleig}(2005)}]{fleig-05}%
  \BibitemOpen
  \bibfield  {author} {\bibinfo {author} {\bibfnamefont {Timo}\ \bibnamefont
  {Fleig}},\ }\bibfield  {title} {\enquote {\bibinfo {title}
  {Spin-orbit-resolved static polarizabilities of group-13 atoms: Four-
  component relativistic configuration interaction and coupled cluster
  calculations},}\ }\href {\doibase 10.1103/PhysRevA.72.052506} {\bibfield
  {journal} {\bibinfo  {journal} {Phys. Rev. A}\ }\textbf {\bibinfo {volume}
  {72}},\ \bibinfo {pages} {052506} (\bibinfo {year} {2005})}\BibitemShut
  {NoStop}%
\bibitem [{\citenamefont {Borschevsky}\ \emph {et~al.}(2012)\citenamefont
  {Borschevsky}, \citenamefont {Zelovich}, \citenamefont {Eliav},\ and\
  \citenamefont {Kaldor}}]{borschevsky-12}%
  \BibitemOpen
  \bibfield  {author} {\bibinfo {author} {\bibfnamefont {Anastasia}\
  \bibnamefont {Borschevsky}}, \bibinfo {author} {\bibfnamefont {Tamar}\
  \bibnamefont {Zelovich}}, \bibinfo {author} {\bibfnamefont {Ephraim}\
  \bibnamefont {Eliav}}, \ and\ \bibinfo {author} {\bibfnamefont {Uzi}\
  \bibnamefont {Kaldor}},\ }\bibfield  {title} {\enquote {\bibinfo {title}
  {Precision of calculated static polarizabilities: {Ga, In and Tl atoms}},}\
  }\href {\doibase https://doi.org/10.1016/j.chemphys.2011.05.011} {\bibfield
  {journal} {\bibinfo  {journal} {Chemical Physics}\ }\textbf {\bibinfo
  {volume} {395}},\ \bibinfo {pages} {104 -- 107} (\bibinfo {year} {2012})},\
  \bibinfo {note} {recent Advances and Applications of Relativistic Quantum
  Chemistry}\BibitemShut {NoStop}%
\bibitem [{\citenamefont {Buchachenko}(2010)}]{buchachenko-10}%
  \BibitemOpen
  \bibfield  {author} {\bibinfo {author} {\bibfnamefont {A.~A.}\ \bibnamefont
  {Buchachenko}},\ }\bibfield  {title} {\enquote {\bibinfo {title}
  {State-interacting spin-orbit configuration interaction method for {J}-
  resolved anisotropic static dipole polarizabilities: Application to {Al, Ga,
  In, and Tl atoms}},}\ }\href {\doibase 10.1134/S0036024410130200} {\bibfield
  {journal} {\bibinfo  {journal} {Russian Journal of Physical Chemistry A}\
  }\textbf {\bibinfo {volume} {84}},\ \bibinfo {pages} {2325--2333} (\bibinfo
  {year} {2010})}\BibitemShut {NoStop}%
\bibitem [{\citenamefont {Lupinetti}\ and\ \citenamefont
  {Thakkar}(2005)}]{lupinetti-05}%
  \BibitemOpen
  \bibfield  {author} {\bibinfo {author} {\bibfnamefont {Concetta}\
  \bibnamefont {Lupinetti}}\ and\ \bibinfo {author} {\bibfnamefont {Ajit~J.}\
  \bibnamefont {Thakkar}},\ }\bibfield  {title} {\enquote {\bibinfo {title}
  {Polarizabilities and hyperpolarizabilities for the atoms {Al, Si, P, S, Cl,
  and Ar}: Coupled cluster calculations},}\ }\href {\doibase 10.1063/1.1834512}
  {\bibfield  {journal} {\bibinfo  {journal} {The Journal of Chemical Physics}\
  }\textbf {\bibinfo {volume} {122}},\ \bibinfo {pages} {044301} (\bibinfo
  {year} {2005})},\ \Eprint
  {http://arxiv.org/abs/https://doi.org/10.1063/1.1834512}
  {https://doi.org/10.1063/1.1834512} \BibitemShut {NoStop}%
\bibitem [{\citenamefont {Fuentealba}(2004)}]{fuentealba-04}%
  \BibitemOpen
  \bibfield  {author} {\bibinfo {author} {\bibfnamefont {P.}~\bibnamefont
  {Fuentealba}},\ }\bibfield  {title} {\enquote {\bibinfo {title} {The static
  dipole polarizability of aluminium atom: discrepancy between theory and
  experiment},}\ }\href {\doibase https://doi.org/10.1016/j.cplett.2004.09.013}
  {\bibfield  {journal} {\bibinfo  {journal} {Chemical Physics Letters}\
  }\textbf {\bibinfo {volume} {397}},\ \bibinfo {pages} {459 -- 461} (\bibinfo
  {year} {2004})}\BibitemShut {NoStop}%
\bibitem [{\citenamefont {Chu}\ and\ \citenamefont {Dalgarno}(2004)}]{chu-04}%
  \BibitemOpen
  \bibfield  {author} {\bibinfo {author} {\bibfnamefont {X.}~\bibnamefont
  {Chu}}\ and\ \bibinfo {author} {\bibfnamefont {A.}~\bibnamefont {Dalgarno}},\
  }\bibfield  {title} {\enquote {\bibinfo {title} {Linear response
  time-dependent density functional theory for van der waals coefficients},}\
  }\href {\doibase 10.1063/1.1779576} {\bibfield  {journal} {\bibinfo
  {journal} {The Journal of Chemical Physics}\ }\textbf {\bibinfo {volume}
  {121}},\ \bibinfo {pages} {4083--4088} (\bibinfo {year} {2004})},\ \Eprint
  {http://arxiv.org/abs/https://doi.org/10.1063/1.1779576}
  {https://doi.org/10.1063/1.1779576} \BibitemShut {NoStop}%
\bibitem [{\citenamefont {Milani}\ \emph {et~al.}(1990)\citenamefont {Milani},
  \citenamefont {Moullet},\ and\ \citenamefont {de~Heer}}]{milani-90}%
  \BibitemOpen
  \bibfield  {author} {\bibinfo {author} {\bibfnamefont {Paolo}\ \bibnamefont
  {Milani}}, \bibinfo {author} {\bibfnamefont {I.}~\bibnamefont {Moullet}}, \
  and\ \bibinfo {author} {\bibfnamefont {Walt~A.}\ \bibnamefont {de~Heer}},\
  }\bibfield  {title} {\enquote {\bibinfo {title} {Experimental and theoretical
  electric dipole polarizabilities of {Al and Al$_2$}},}\ }\href {\doibase
  10.1103/PhysRevA.42.5150} {\bibfield  {journal} {\bibinfo  {journal} {Phys.
  Rev. A}\ }\textbf {\bibinfo {volume} {42}},\ \bibinfo {pages} {5150--5154}
  (\bibinfo {year} {1990})}\BibitemShut {NoStop}%
\bibitem [{\citenamefont {Sarkisov}\ \emph {et~al.}(2006)\citenamefont
  {Sarkisov}, \citenamefont {Beigman}, \citenamefont {Shevelko},\ and\
  \citenamefont {Struve}}]{sarkisov-06}%
  \BibitemOpen
  \bibfield  {author} {\bibinfo {author} {\bibfnamefont {G.~S.}\ \bibnamefont
  {Sarkisov}}, \bibinfo {author} {\bibfnamefont {I.~L.}\ \bibnamefont
  {Beigman}}, \bibinfo {author} {\bibfnamefont {V.~P.}\ \bibnamefont
  {Shevelko}}, \ and\ \bibinfo {author} {\bibfnamefont {K.~W.}\ \bibnamefont
  {Struve}},\ }\bibfield  {title} {\enquote {\bibinfo {title} {Interferometric
  measurements of dynamic polarizabilities for metal atoms using electrically
  exploding wires in vacuum},}\ }\href {\doibase 10.1103/PhysRevA.73.042501}
  {\bibfield  {journal} {\bibinfo  {journal} {Phys. Rev. A}\ }\textbf {\bibinfo
  {volume} {73}},\ \bibinfo {pages} {042501} (\bibinfo {year}
  {2006})}\BibitemShut {NoStop}%
\bibitem [{\citenamefont {Guella}\ \emph {et~al.}(1984)\citenamefont {Guella},
  \citenamefont {Miller}, \citenamefont {Bederson}, \citenamefont {Stockdale},\
  and\ \citenamefont {Jaduszliwer}}]{guella-84}%
  \BibitemOpen
  \bibfield  {author} {\bibinfo {author} {\bibfnamefont {T.~P.}\ \bibnamefont
  {Guella}}, \bibinfo {author} {\bibfnamefont {Thomas~M.}\ \bibnamefont
  {Miller}}, \bibinfo {author} {\bibfnamefont {B.}~\bibnamefont {Bederson}},
  \bibinfo {author} {\bibfnamefont {J.~A.~D.}\ \bibnamefont {Stockdale}}, \
  and\ \bibinfo {author} {\bibfnamefont {B.}~\bibnamefont {Jaduszliwer}},\
  }\bibfield  {title} {\enquote {\bibinfo {title} {Polarizability of
  $5{s}^{2}5p(^{2}p_{\frac{1}{2}})$ atomic indium},}\ }\href {\doibase
  10.1103/PhysRevA.29.2977} {\bibfield  {journal} {\bibinfo  {journal} {Phys.
  Rev. A}\ }\textbf {\bibinfo {volume} {29}},\ \bibinfo {pages} {2977--2980}
  (\bibinfo {year} {1984})}\BibitemShut {NoStop}%
\bibitem [{\citenamefont {Ma}\ \emph {et~al.}(2015)\citenamefont {Ma},
  \citenamefont {Indergaard}, \citenamefont {Zhang}, \citenamefont {Larkin},
  \citenamefont {Moro},\ and\ \citenamefont {de~Heer}}]{lei-15}%
  \BibitemOpen
  \bibfield  {author} {\bibinfo {author} {\bibfnamefont {Lei}\ \bibnamefont
  {Ma}}, \bibinfo {author} {\bibfnamefont {John}\ \bibnamefont {Indergaard}},
  \bibinfo {author} {\bibfnamefont {Baiqian}\ \bibnamefont {Zhang}}, \bibinfo
  {author} {\bibfnamefont {Ilia}\ \bibnamefont {Larkin}}, \bibinfo {author}
  {\bibfnamefont {Ramiro}\ \bibnamefont {Moro}}, \ and\ \bibinfo {author}
  {\bibfnamefont {Walt~A.}\ \bibnamefont {de~Heer}},\ }\bibfield  {title}
  {\enquote {\bibinfo {title} {Measured atomic ground-state polarizabilities of
  35 metallic elements},}\ }\href {\doibase 10.1103/PhysRevA.91.010501}
  {\bibfield  {journal} {\bibinfo  {journal} {Phys. Rev. A}\ }\textbf {\bibinfo
  {volume} {91}},\ \bibinfo {pages} {010501} (\bibinfo {year}
  {2015})}\BibitemShut {NoStop}%
\bibitem [{\citenamefont {Mani}\ \emph {et~al.}(2017)\citenamefont {Mani},
  \citenamefont {Chattopadhyay},\ and\ \citenamefont {Angom}}]{mani-17}%
  \BibitemOpen
  \bibfield  {author} {\bibinfo {author} {\bibfnamefont {B.K.}\ \bibnamefont
  {Mani}}, \bibinfo {author} {\bibfnamefont {S.}~\bibnamefont {Chattopadhyay}},
  \ and\ \bibinfo {author} {\bibfnamefont {D.}~\bibnamefont {Angom}},\
  }\bibfield  {title} {\enquote {\bibinfo {title} {Rccpac: A parallel
  relativistic coupled-cluster program for closed-shell and one-valence atoms
  and ions in fortran},}\ }\href {\doibase
  https://doi.org/10.1016/j.cpc.2016.11.008} {\bibfield  {journal} {\bibinfo
  {journal} {Computer Physics Communications}\ }\textbf {\bibinfo {volume}
  {213}},\ \bibinfo {pages} {136 -- 154} (\bibinfo {year} {2017})}\BibitemShut
  {NoStop}%
\bibitem [{\citenamefont {Kumar}\ \emph {et~al.}(2020)\citenamefont {Kumar},
  \citenamefont {Chattopadhyay}, \citenamefont {Mani},\ and\ \citenamefont
  {Angom}}]{ravi-20}%
  \BibitemOpen
  \bibfield  {author} {\bibinfo {author} {\bibfnamefont {Ravi}\ \bibnamefont
  {Kumar}}, \bibinfo {author} {\bibfnamefont {S.}~\bibnamefont
  {Chattopadhyay}}, \bibinfo {author} {\bibfnamefont {B.~K.}\ \bibnamefont
  {Mani}}, \ and\ \bibinfo {author} {\bibfnamefont {D.}~\bibnamefont {Angom}},\
  }\bibfield  {title} {\enquote {\bibinfo {title} {Electric dipole
  polarizability of group-13 ions using perturbed relativistic coupled-cluster
  theory: Importance of nonlinear terms},}\ }\href {\doibase
  10.1103/PhysRevA.101.012503} {\bibfield  {journal} {\bibinfo  {journal}
  {Phys. Rev. A}\ }\textbf {\bibinfo {volume} {101}},\ \bibinfo {pages}
  {012503} (\bibinfo {year} {2020})}\BibitemShut {NoStop}%
\bibitem [{\citenamefont {Kumar}\ \emph {et~al.}(2021)\citenamefont {Kumar},
  \citenamefont {Chattopadhyay}, \citenamefont {Angom},\ and\ \citenamefont
  {Mani}}]{ravi-21}%
  \BibitemOpen
  \bibfield  {author} {\bibinfo {author} {\bibfnamefont {Ravi}\ \bibnamefont
  {Kumar}}, \bibinfo {author} {\bibfnamefont {S.}~\bibnamefont
  {Chattopadhyay}}, \bibinfo {author} {\bibfnamefont {D.}~\bibnamefont
  {Angom}}, \ and\ \bibinfo {author} {\bibfnamefont {B.~K.}\ \bibnamefont
  {Mani}},\ }\bibfield  {title} {\enquote {\bibinfo {title} {Fock-space
  relativistic coupled-cluster calculation of a hyperfine-induced
  ${}^{1}{S}_{0}\ensuremath{\rightarrow}{}^{3}{P}_{0}^{o}$ clock transition in
  ${\mathrm{al}}^{+}$},}\ }\href {\doibase 10.1103/PhysRevA.103.022801}
  {\bibfield  {journal} {\bibinfo  {journal} {Phys. Rev. A}\ }\textbf {\bibinfo
  {volume} {103}},\ \bibinfo {pages} {022801} (\bibinfo {year}
  {2021})}\BibitemShut {NoStop}%
\bibitem [{\citenamefont {Safronova}\ \emph {et~al.}(1999)\citenamefont
  {Safronova}, \citenamefont {Johnson},\ and\ \citenamefont
  {Derevianko}}]{safronova-99}%
  \BibitemOpen
  \bibfield  {author} {\bibinfo {author} {\bibfnamefont {M.~S.}\ \bibnamefont
  {Safronova}}, \bibinfo {author} {\bibfnamefont {W.~R.}\ \bibnamefont
  {Johnson}}, \ and\ \bibinfo {author} {\bibfnamefont {A.}~\bibnamefont
  {Derevianko}},\ }\bibfield  {title} {\enquote {\bibinfo {title} {Relativistic
  many-body calculations of energy levels, hyperfine constants, electric-dipole
  matrix elements, and static polarizabilities for alkali-metal atoms},}\
  }\href {\doibase 10.1103/PhysRevA.60.4476} {\bibfield  {journal} {\bibinfo
  {journal} {Phys. Rev. A}\ }\textbf {\bibinfo {volume} {60}},\ \bibinfo
  {pages} {4476--4487} (\bibinfo {year} {1999})}\BibitemShut {NoStop}%
\bibitem [{\citenamefont {Derevianko}\ \emph {et~al.}(1999)\citenamefont
  {Derevianko}, \citenamefont {Johnson}, \citenamefont {Safronova},\ and\
  \citenamefont {Babb}}]{derevianko-99}%
  \BibitemOpen
  \bibfield  {author} {\bibinfo {author} {\bibfnamefont {A.}~\bibnamefont
  {Derevianko}}, \bibinfo {author} {\bibfnamefont {W.~R.}\ \bibnamefont
  {Johnson}}, \bibinfo {author} {\bibfnamefont {M.~S.}\ \bibnamefont
  {Safronova}}, \ and\ \bibinfo {author} {\bibfnamefont {J.~F.}\ \bibnamefont
  {Babb}},\ }\bibfield  {title} {\enquote {\bibinfo {title} {High-precision
  calculations of dispersion coefficients, static dipole polarizabilities, and
  atom-wall interaction constants for alkali-metal atoms},}\ }\href {\doibase
  10.1103/PhysRevLett.82.3589} {\bibfield  {journal} {\bibinfo  {journal}
  {Phys. Rev. Lett.}\ }\textbf {\bibinfo {volume} {82}},\ \bibinfo {pages}
  {3589--3592} (\bibinfo {year} {1999})}\BibitemShut {NoStop}%
\bibitem [{\citenamefont {Mitroy}\ \emph {et~al.}(2010)\citenamefont {Mitroy},
  \citenamefont {Safronova},\ and\ \citenamefont {Clark}}]{mitroy-10}%
  \BibitemOpen
  \bibfield  {author} {\bibinfo {author} {\bibfnamefont {J}~\bibnamefont
  {Mitroy}}, \bibinfo {author} {\bibfnamefont {M~S}\ \bibnamefont {Safronova}},
  \ and\ \bibinfo {author} {\bibfnamefont {Charles~W}\ \bibnamefont {Clark}},\
  }\bibfield  {title} {\enquote {\bibinfo {title} {Theory and applications of
  atomic and ionic polarizabilities},}\ }\href {\doibase
  10.1088/0953-4075/43/20/202001} {\bibfield  {journal} {\bibinfo  {journal}
  {Journal of Physics B: Atomic, Molecular and Optical Physics}\ }\textbf
  {\bibinfo {volume} {43}},\ \bibinfo {pages} {202001} (\bibinfo {year}
  {2010})}\BibitemShut {NoStop}%
\bibitem [{\citenamefont {Schwerdtfeger}\ and\ \citenamefont
  {Nagle}(2019)}]{peter-19}%
  \BibitemOpen
  \bibfield  {author} {\bibinfo {author} {\bibfnamefont {Peter}\ \bibnamefont
  {Schwerdtfeger}}\ and\ \bibinfo {author} {\bibfnamefont {Jeffrey~K.}\
  \bibnamefont {Nagle}},\ }\bibfield  {title} {\enquote {\bibinfo {title} {2018
  table of static dipole polarizabilities of the neutral elements in the
  periodic table},}\ }\href {\doibase 10.1080/00268976.2018.1535143} {\bibfield
   {journal} {\bibinfo  {journal} {Molecular Physics}\ }\textbf {\bibinfo
  {volume} {117}},\ \bibinfo {pages} {1200--1225} (\bibinfo {year} {2019})},\
  \Eprint {http://arxiv.org/abs/https://doi.org/10.1080/00268976.2018.1535143}
  {https://doi.org/10.1080/00268976.2018.1535143} \BibitemShut {NoStop}%
\bibitem [{\citenamefont {Mohanty}\ \emph {et~al.}(1991)\citenamefont
  {Mohanty}, \citenamefont {Parpia},\ and\ \citenamefont
  {Clementi}}]{mohanty-91}%
  \BibitemOpen
  \bibfield  {author} {\bibinfo {author} {\bibfnamefont {A.~K.}\ \bibnamefont
  {Mohanty}}, \bibinfo {author} {\bibfnamefont {F.~A.}\ \bibnamefont {Parpia}},
  \ and\ \bibinfo {author} {\bibfnamefont {E.}~\bibnamefont {Clementi}},\
  }\bibfield  {title} {\enquote {\bibinfo {title} {Kinetically balanced
  geometric gaussian basis set calculations for relativistic many-electron
  atoms},}\ }in\ \href@noop {} {\emph {\bibinfo {booktitle} {Modern Techniques
  in Computational Chemistry: MOTECC-91}}},\ \bibinfo {editor} {edited by\
  \bibinfo {editor} {\bibfnamefont {E.}~\bibnamefont {Clementi}}}\ (\bibinfo
  {publisher} {ESCOM},\ \bibinfo {year} {1991})\BibitemShut {NoStop}%
\bibitem [{\citenamefont {Stanton}\ and\ \citenamefont
  {Havriliak}(1984)}]{stanton-84}%
  \BibitemOpen
  \bibfield  {author} {\bibinfo {author} {\bibfnamefont {Richard~E.}\
  \bibnamefont {Stanton}}\ and\ \bibinfo {author} {\bibfnamefont {Stephen}\
  \bibnamefont {Havriliak}},\ }\bibfield  {title} {\enquote {\bibinfo {title}
  {Kinetic balance: A partial solution to the problem of variational safety in
  dirac calculations},}\ }\href {\doibase 10.1063/1.447865} {\bibfield
  {journal} {\bibinfo  {journal} {J. Chem. Phys.}\ }\textbf {\bibinfo {volume}
  {81}},\ \bibinfo {pages} {1910--1918} (\bibinfo {year} {1984})}\BibitemShut
  {NoStop}%
\bibitem [{\citenamefont {Mani}\ \emph {et~al.}(2009)\citenamefont {Mani},
  \citenamefont {Latha},\ and\ \citenamefont {Angom}}]{mani-09}%
  \BibitemOpen
  \bibfield  {author} {\bibinfo {author} {\bibfnamefont {B.~K.}\ \bibnamefont
  {Mani}}, \bibinfo {author} {\bibfnamefont {K.~V.~P.}\ \bibnamefont {Latha}},
  \ and\ \bibinfo {author} {\bibfnamefont {D.}~\bibnamefont {Angom}},\
  }\bibfield  {title} {\enquote {\bibinfo {title} {Relativistic coupled-cluster
  calculations of $^{20}\text{N}\text{e}$, $^{40}\text{A}\text{r}$,
  $^{84}\text{K}\text{r}$, and $^{129}\text{X}\text{e}$: Correlation energies
  and dipole polarizabilities},}\ }\href {\doibase 10.1103/PhysRevA.80.062505}
  {\bibfield  {journal} {\bibinfo  {journal} {Phys. Rev. A}\ }\textbf {\bibinfo
  {volume} {80}},\ \bibinfo {pages} {062505} (\bibinfo {year}
  {2009})}\BibitemShut {NoStop}%
\bibitem [{\citenamefont {Mani}\ and\ \citenamefont {Angom}(2010)}]{mani-10}%
  \BibitemOpen
  \bibfield  {author} {\bibinfo {author} {\bibfnamefont {B.~K.}\ \bibnamefont
  {Mani}}\ and\ \bibinfo {author} {\bibfnamefont {D.}~\bibnamefont {Angom}},\
  }\bibfield  {title} {\enquote {\bibinfo {title} {Atomic properties calculated
  by relativistic coupled-cluster theory without truncation: Hyperfine
  constants of {Mg}$^{+}$, {Ca}$^{+}$, {Sr}$^{+}$, and {Ba}$^{+}$},}\ }\href
  {\doibase 10.1103/PhysRevA.81.042514} {\bibfield  {journal} {\bibinfo
  {journal} {Phys. Rev. A}\ }\textbf {\bibinfo {volume} {81}},\ \bibinfo
  {pages} {042514} (\bibinfo {year} {2010})}\BibitemShut {NoStop}%
\bibitem [{\citenamefont {Chattopadhyay}\ \emph {et~al.}(2012)\citenamefont
  {Chattopadhyay}, \citenamefont {Mani},\ and\ \citenamefont
  {Angom}}]{chattopadhyay-12b}%
  \BibitemOpen
  \bibfield  {author} {\bibinfo {author} {\bibfnamefont {S.}~\bibnamefont
  {Chattopadhyay}}, \bibinfo {author} {\bibfnamefont {B.~K.}\ \bibnamefont
  {Mani}}, \ and\ \bibinfo {author} {\bibfnamefont {D.}~\bibnamefont {Angom}},\
  }\bibfield  {title} {\enquote {\bibinfo {title} {Perturbed coupled-cluster
  theory to calculate dipole polarizabilities of closed-shell systems:
  Application to {Ar, Kr, Xe, and Rn}},}\ }\href {\doibase
  10.1103/PhysRevA.86.062508} {\bibfield  {journal} {\bibinfo  {journal} {Phys.
  Rev. A}\ }\textbf {\bibinfo {volume} {86}},\ \bibinfo {pages} {062508}
  (\bibinfo {year} {2012})}\BibitemShut {NoStop}%
\bibitem [{\citenamefont {Pulay}(1980)}]{pulay-80}%
  \BibitemOpen
  \bibfield  {author} {\bibinfo {author} {\bibfnamefont {Péter}\ \bibnamefont
  {Pulay}},\ }\bibfield  {title} {\enquote {\bibinfo {title} {Convergence
  acceleration of iterative sequences. the case of scf iteration},}\ }\href
  {\doibase 10.1016/0009-2614(80)80396-4} {\bibfield  {journal} {\bibinfo
  {journal} {Chem. Phys. Lett.}\ }\textbf {\bibinfo {volume} {73}},\ \bibinfo
  {pages} {393 -- 398} (\bibinfo {year} {1980})}\BibitemShut {NoStop}%
\bibitem [{\citenamefont {{Fritzsche}}\ \emph {et~al.}(2008)\citenamefont
  {{Fritzsche}}, \citenamefont {{Mani}},\ and\ \citenamefont
  {{Angom}}}]{fritzsche-08}%
  \BibitemOpen
  \bibfield  {author} {\bibinfo {author} {\bibfnamefont {Stephan}\ \bibnamefont
  {{Fritzsche}}}, \bibinfo {author} {\bibfnamefont {Brajesh~K.}\ \bibnamefont
  {{Mani}}}, \ and\ \bibinfo {author} {\bibfnamefont {Dilip}\ \bibnamefont
  {{Angom}}},\ }\bibfield  {title} {\enquote {\bibinfo {title} {{Chapter 10 A
  Computer-Algebraic Approach to the Derivation of Feynman-Goldstone
  Perturbation Expansions for Open-Shell Atoms and Molecules}},}\ }\href
  {\doibase 10.1016/S0065-3276(07)53010-8} {\bibfield  {journal} {\bibinfo
  {journal} {Advances in Quantum Chemistry}\ }\textbf {\bibinfo {volume}
  {53}},\ \bibinfo {pages} {177--215} (\bibinfo {year} {2008})}\BibitemShut
  {NoStop}%
\bibitem [{\citenamefont {Lindgren}\ and\ \citenamefont {Morrison}(2nd Edition,
  1986)}]{lindgren-86}%
  \BibitemOpen
  \bibfield  {author} {\bibinfo {author} {\bibfnamefont {I.}~\bibnamefont
  {Lindgren}}\ and\ \bibinfo {author} {\bibfnamefont {J.}~\bibnamefont
  {Morrison}},\ }\href@noop {} {\emph {\bibinfo {title} {Atomic Many-Body
  Theory}}}\ (\bibinfo  {publisher} {Springer},\ \bibinfo {address} {Berlin},\
  \bibinfo {year} {2nd Edition, 1986})\BibitemShut {NoStop}%
\bibitem [{\citenamefont {J{\"{o}}nsson}\ \emph {et~al.}(2013)\citenamefont
  {J{\"{o}}nsson}, \citenamefont {Gaigalas}, \citenamefont {Biero{\'{n}}},
  \citenamefont {Froese~Fischer},\ and\ \citenamefont {Grant}}]{jonsson-13}%
  \BibitemOpen
  \bibfield  {author} {\bibinfo {author} {\bibfnamefont {P.}~\bibnamefont
  {J{\"{o}}nsson}}, \bibinfo {author} {\bibfnamefont {G.}~\bibnamefont
  {Gaigalas}}, \bibinfo {author} {\bibfnamefont {J.}~\bibnamefont
  {Biero{\'{n}}}}, \bibinfo {author} {\bibfnamefont {C.}~\bibnamefont
  {Froese~Fischer}}, \ and\ \bibinfo {author} {\bibfnamefont {I.~P.}\
  \bibnamefont {Grant}},\ }\bibfield  {title} {\enquote {\bibinfo {title} {New
  version: Grasp2k relativistic atomic structure package},}\ }\href {\doibase
  http://dx.doi.org/10.1016/j.cpc.2013.02.016} {\bibfield  {journal} {\bibinfo
  {journal} {Comp. Phys. Comm.}\ }\textbf {\bibinfo {volume} {184}},\ \bibinfo
  {pages} {2197 -- 2203} (\bibinfo {year} {2013})}\BibitemShut {NoStop}%
\bibitem [{\citenamefont {Grant}\ and\ \citenamefont
  {McKenzie}(1980)}]{grant-80}%
  \BibitemOpen
  \bibfield  {author} {\bibinfo {author} {\bibfnamefont {I~P}\ \bibnamefont
  {Grant}}\ and\ \bibinfo {author} {\bibfnamefont {B~J}\ \bibnamefont
  {McKenzie}},\ }\bibfield  {title} {\enquote {\bibinfo {title} {The transverse
  electron-electron interaction in atomic structure calculations},}\ }\href
  {http://stacks.iop.org/0022-3700/13/i=14/a=007} {\bibfield  {journal}
  {\bibinfo  {journal} {J. Phys. B}\ }\textbf {\bibinfo {volume} {13}},\
  \bibinfo {pages} {2671} (\bibinfo {year} {1980})}\BibitemShut {NoStop}%
\bibitem [{\citenamefont {Uehling}(1935)}]{uehling-35}%
  \BibitemOpen
  \bibfield  {author} {\bibinfo {author} {\bibfnamefont {E.~A.}\ \bibnamefont
  {Uehling}},\ }\bibfield  {title} {\enquote {\bibinfo {title} {Polarization
  effects in the positron theory},}\ }\href {\doibase 10.1103/PhysRev.48.55}
  {\bibfield  {journal} {\bibinfo  {journal} {Phys. Rev.}\ }\textbf {\bibinfo
  {volume} {48}},\ \bibinfo {pages} {55--63} (\bibinfo {year}
  {1935})}\BibitemShut {NoStop}%
\bibitem [{\citenamefont {Fullerton}\ and\ \citenamefont
  {Rinker}(1976)}]{fullerton-76}%
  \BibitemOpen
  \bibfield  {author} {\bibinfo {author} {\bibfnamefont {L.~Wayne}\
  \bibnamefont {Fullerton}}\ and\ \bibinfo {author} {\bibfnamefont {G.~A.}\
  \bibnamefont {Rinker}},\ }\bibfield  {title} {\enquote {\bibinfo {title}
  {Accurate and efficient methods for the evaluation of vacuum-polarization
  potentials of order $z\ensuremath{\alpha}$ and
  $z{\ensuremath{\alpha}}^{2}$},}\ }\href {\doibase 10.1103/PhysRevA.13.1283}
  {\bibfield  {journal} {\bibinfo  {journal} {Phys. Rev. A}\ }\textbf {\bibinfo
  {volume} {13}},\ \bibinfo {pages} {1283--1287} (\bibinfo {year}
  {1976})}\BibitemShut {NoStop}%
\bibitem [{\citenamefont {Shabaev}\ \emph {et~al.}(2013)\citenamefont
  {Shabaev}, \citenamefont {Tupitsyn},\ and\ \citenamefont
  {Yerokhin}}]{shabaev-13}%
  \BibitemOpen
  \bibfield  {author} {\bibinfo {author} {\bibfnamefont {V.~M.}\ \bibnamefont
  {Shabaev}}, \bibinfo {author} {\bibfnamefont {I.~I.}\ \bibnamefont
  {Tupitsyn}}, \ and\ \bibinfo {author} {\bibfnamefont {V.~A.}\ \bibnamefont
  {Yerokhin}},\ }\bibfield  {title} {\enquote {\bibinfo {title} {Model operator
  approach to the lamb shift calculations in relativistic many-electron
  atoms},}\ }\href {\doibase 10.1103/PhysRevA.88.012513} {\bibfield  {journal}
  {\bibinfo  {journal} {Phys. Rev. A}\ }\textbf {\bibinfo {volume} {88}},\
  \bibinfo {pages} {012513} (\bibinfo {year} {2013})}\BibitemShut {NoStop}%
\bibitem [{\citenamefont {Shabaev}\ \emph {et~al.}(2015)\citenamefont
  {Shabaev}, \citenamefont {Tupitsyn},\ and\ \citenamefont
  {Yerokhin}}]{shabaev-15}%
  \BibitemOpen
  \bibfield  {author} {\bibinfo {author} {\bibfnamefont {V.M.}\ \bibnamefont
  {Shabaev}}, \bibinfo {author} {\bibfnamefont {I.I.}\ \bibnamefont
  {Tupitsyn}}, \ and\ \bibinfo {author} {\bibfnamefont {V.A.}\ \bibnamefont
  {Yerokhin}},\ }\bibfield  {title} {\enquote {\bibinfo {title} {Qedmod:
  Fortran program for calculating the model lamb-shift operator},}\ }\href
  {\doibase https://doi.org/10.1016/j.cpc.2014.12.002} {\bibfield  {journal}
  {\bibinfo  {journal} {Computer Physics Communications}\ }\textbf {\bibinfo
  {volume} {189}},\ \bibinfo {pages} {175--181} (\bibinfo {year}
  {2015})}\BibitemShut {NoStop}%
\bibitem [{\citenamefont {Grant}(2006)}]{grant-06}%
  \BibitemOpen
  \bibfield  {author} {\bibinfo {author} {\bibfnamefont {Ian}\ \bibnamefont
  {Grant}},\ }\bibfield  {title} {\enquote {\bibinfo {title} {Relativistic
  atomic structure},}\ }in\ \href {\doibase 10.1007/978-0-387-26308-3_22}
  {\emph {\bibinfo {booktitle} {Springer Handbook of Atomic, Molecular, and
  Optical Physics}}},\ \bibinfo {editor} {edited by\ \bibinfo {editor}
  {\bibfnamefont {Gordon}\ \bibnamefont {Drake}}}\ (\bibinfo  {publisher}
  {Springer},\ \bibinfo {address} {New York},\ \bibinfo {year} {2006})\ pp.\
  \bibinfo {pages} {325--357}\BibitemShut {NoStop}%
\bibitem [{nis(2013)}]{nist}%
  \BibitemOpen
  \href@noop {} {\enquote {\bibinfo {title} {Nist atomic spectroscopic
  database},}\ }\bibinfo {howpublished}
  {\url{https://physics.nist.gov/PhysRefData/ASD/levels_form.html}} (\bibinfo
  {year} {2013})\BibitemShut {NoStop}%
\bibitem [{\citenamefont {Safronova}\ \emph {et~al.}(2013)\citenamefont
  {Safronova}, \citenamefont {Safronova},\ and\ \citenamefont
  {Porsev}}]{safronova-13b}%
  \BibitemOpen
  \bibfield  {author} {\bibinfo {author} {\bibfnamefont {M.~S.}\ \bibnamefont
  {Safronova}}, \bibinfo {author} {\bibfnamefont {U.~I.}\ \bibnamefont
  {Safronova}}, \ and\ \bibinfo {author} {\bibfnamefont {S.~G.}\ \bibnamefont
  {Porsev}},\ }\bibfield  {title} {\enquote {\bibinfo {title}
  {{Polarizabilities, Stark shifts, and lifetimes of the In atom}},}\ }\href
  {\doibase 10.1103/PhysRevA.87.032513} {\bibfield  {journal} {\bibinfo
  {journal} {Phys. Rev. A}\ }\textbf {\bibinfo {volume} {87}},\ \bibinfo
  {pages} {032513} (\bibinfo {year} {2013})}\BibitemShut {NoStop}%
\bibitem [{\citenamefont {Chattopadhyay}\ \emph {et~al.}(2014)\citenamefont
  {Chattopadhyay}, \citenamefont {Mani},\ and\ \citenamefont
  {Angom}}]{chattopadhyay-14}%
  \BibitemOpen
  \bibfield  {author} {\bibinfo {author} {\bibfnamefont {S.}~\bibnamefont
  {Chattopadhyay}}, \bibinfo {author} {\bibfnamefont {B.~K.}\ \bibnamefont
  {Mani}}, \ and\ \bibinfo {author} {\bibfnamefont {D.}~\bibnamefont {Angom}},\
  }\bibfield  {title} {\enquote {\bibinfo {title} {Electric dipole
  polarizability of alkaline-earth-metal atoms from perturbed relativistic
  coupled-cluster theory with triples},}\ }\href {\doibase
  10.1103/PhysRevA.89.022506} {\bibfield  {journal} {\bibinfo  {journal} {Phys.
  Rev. A}\ }\textbf {\bibinfo {volume} {89}},\ \bibinfo {pages} {022506}
  (\bibinfo {year} {2014})}\BibitemShut {NoStop}%
\end{thebibliography}%

\end{document}